\documentclass[preprint,12pt, 3p]{elsarticle}

\usepackage{amsmath}
\usepackage{amssymb}
\usepackage{amsthm}
\usepackage{hyperref}
\usepackage[T1]{fontenc}
\usepackage{soul}

\usepackage{xcolor}

\newcommand{\sbody}[2]{{\textstyle\frac{#1}{#2}}}

\newcommand{\be}{\begin{equation}} 
\newcommand{\ee}{\end{equation}} 
\newcommand{\bea}{\begin{eqnarray}} 
\newcommand{\eea}{\end{eqnarray}}

\def\slash#1{#1\!\!\!\raise.15ex\hbox {/}}
\newcommand{\slD}{\,\raise.15ex\hbox{$/$}\kern-.27em\hbox{$\!\!\!D$}}
\newcommand{\slpartial}{\raise.15ex\hbox{$/$}\kern-.57em\hbox{$\partial$}}

\def\2int{\int_{0}^{1}  \!\!\! du \! \int_{0}^{1}  \!\!\! dv}

\def\e{\mbox{e}}

\def\Eins{{\mathchoice {\rm 1\mskip-4mu l} {\rm 1\mskip-4mu l}
{\rm 1\mskip-4.5mu l} {\rm 1\mskip-5mu l}}}
\def\Z{{\mathchoice {\hbox{$\sf\textstyle Z\kern-0.4em Z$}}
{\hbox{$\sf\textstyle Z\kern-0.4em Z$}}
{\hbox{$\sf\scriptstyle Z\kern-0.3em Z$}}
{\hbox{$\sf\scriptscriptstyle Z\kern-0.2em Z$}}}}

\def\no{\noindent}

\def\Eins{{\mathchoice {\rm 1\mskip-4mu l} {\rm 1\mskip-4mu l}{\rm 1\mskip-4.5mu l} {\rm 1\mskip-5mu l}}}

\def\half{\frac{1}{2}}

\def\fourth{\frac{1}{4}}

\def\ab{{\alpha\beta}}

\def\mn{{\mu\nu}}
\def\abs#1{\left| #1\right|}

\def\Tr{{\rm Tr}\,}
\def\tr{{\rm tr}\,}

\def\e{\,{\rm e}}

\def\b0{{\bf 0}}

\def\la{\langle}
\def\ra{\rangle}
\def\bear{\begin{eqnarray}}
\def\ear{\end{eqnarray}\noindent}
\def\bec{\blue\begin{equation}}
\def\eec{\end{equation}\black\noindent}
\def\bearc{\blue\begin{eqnarray}}
\def\earc{\end{eqnarray}\black\noindent}
\def\benn{\begin{enumerate}}
\def\enn{\end{enumerate}}

\def\intdp3{\int\frac{d^3p}{(2\pi)^3}}
\def\intdp4{\int\frac{d^4p}{(2\pi)^4}}

\newcommand{\Dhalf}{{D\over 2}}

\def\Dhalf{\frac{D}{2}}
\def\4piTD{{(4\pi T)}^{-{D\over 2}}}
\def\4piT4{{(4\pi T)}^{-2}}

\newcommand{\Gd}{{\dot G}}

\newcommand{\slG}{{{\dot G}\!\!\!\! \raise.15ex\hbox {/}}}

\begin{document}

\begin{frontmatter}



\title{Unified worldline treatment of Yukawa and axial couplings
}

             

\author[label1,label2]{F. Bastianelli}

\affiliation[label1]{organization={Dipartimento di Fisica e Astronomia ``Augusto Righi'', Universit\`a di Bologna}, 
addressline={\\Via Irnerio 46}, postcode={I-40126}, 
city={Bologna}, country={Italy\\[2mm]}}

\affiliation[label2]{organization={INFN, Sezione di Bologna}, 
addressline={\\Via Irnerio 46}, postcode={I-40126}, 
city={Bologna}, country={Italy\\[2mm]}}

\author[label3,label2]{O. Corradini}

\affiliation[label3]{organization={Dipartimento di Scienze Fisiche, Informatiche e Matematiche,\\ Universit\`a degli Studi di Modena e Reggio Emilia}, addressline={\\Via Campi 213/A}, postcode={I-41125}, 
city={Modena}, country={Italy\\[2mm]}}

\author[label4,label3]{J.P. Edwards}

\affiliation[label4]{organization={Centre for Mathematical Sciences, University of Plymouth}, city={\\ Plymouth}, postcode={PL4 8AA}, country={UK\\[2mm]}}

\author[label5,label6]{D.G.C. McKeon}

\affiliation[label5]{organization={Department of Applied Mathematics, The University of Western Ontario}
, city={\\ London}, state={Ontario}, postcode={N6A 5B7}, country={Canada\\[2mm]}}

\affiliation[label6]{organization={Algoma University}, city={\\ Sault Ste. Marie}, state={Ontario}, postcode={P6A 2G4}, country={Canada\\[2mm]}}

\author[label7]{C. Schubert}

\affiliation[label7]{organization={Facultad de Ciencias F\'isico-Matem\'aticas,\\ Universidad Michoacana de San Nicol\'as de
Hidalgo}, addressline={\\Avenida Francisco J. M\'ujica}, postcode={58060}, city={Morelia}, state={Michoac\'an}, country={M\'exico}}


\begin{abstract}
We provide a worldline representation of the one-loop effective action for a Dirac particle coupled to external scalar, pseudoscalar, vector and axialvector fields. 
Extending previous work by two of the authors on the pure vector-axialvector case to all four couplings, it allows one to treat the real and the imaginary parts of the effective action in a unified manner, at the price of having a non-Hermitian Hamiltonian. 

Unlike existing worldline representations, our new worldline action contains terms with an odd number of Grassmann fields, leading to ordering problems
that in the worldline formalism are usually encountered only in curved space. Drawing on the highly developed technology for worldline path-integrals in gravity, 
we employ the Time Slicing regularisation of the path integral which comes about with a specific ``counterterm Lagrangian'', which we calculate
once and for all and non-perturbatively, to provide unambiguous rules to treat products of distributions occurring in some diagrams of the one-dimensional worldline theory. 
We then employ the usual worldline machinery to lay out the rules for the calculation of the effective action itself as well as the corresponding one-loop amplitudes.

We test the formalism on the calculation of various heat-kernel coefficients, self energies and scattering amplitudes, including the Higgs decay into two photons or gluons and the PCAC
relation. In all cases we find perfect agreement with more established approaches. 
\end{abstract}
%
%
%

\end{frontmatter}

\tableofcontents
\newpage

\section{Introduction}
\setcounter{equation}{0}
\renewcommand\theequation{\arabic{part}.\arabic{section}.\arabic{equation}}
\numberwithin{equation}{section}

\vskip-3pt

The worldline formalism offers a first quantised alternative to the standard Feynman diagram method in perturbative quantum field theory (QFT). 
Originally suggested by Feynman for multiloop scalar \cite{feynman50} and spinor QED \cite{feynman51} (see also \cite{fradkin,casalbuoni})
it experienced a renaissance in the early nineties in the context of one-loop QCD and string theory amplitudes
through the work of Bern and Kosower \cite{berkosNPB} and Strassler \cite{strassler1}, to some extent anticipated by Polyakov \cite{polyakov-book}
(see also \cite{Bastianelli:1991be,Bastianelli:1992ct,mckeon,mckshe}).
The main idea is to reformulate QFT not in terms of its underlying fields but rather in terms of path integrals for 
the corresponding relativistic point particles. Since then it has found a number of applications in the QFT of scalars and spinors coupled to abelian and non-abelian gauge fields, both in flat space and in the presence of electromagnetic or gravitational backgrounds, and has provided a powerful alternative approach to 
the calculation of effective actions, propagators and scattering amplitudes (see the reviews \cite{41,126,153} and books \cite{Bastianelli:2006rx,book}).

Naturally, soon the need was felt for the construction of worldline representations for the full set of Standard Model couplings. 
In \cite{12,16,dhogag1,dhogag2} such representations were developed for the closed fermion loop coupled to external scalar, pseudoscalar, vector and axialvector
fields, which in Minkowskian field theory would be described by the space-time action\footnote{
In Minkowski space we use the mostly plus metric $\eta_{\mu\nu}$, with gamma matrices satisfying $\{\gamma^\mu,\gamma^\nu\} = -2 \eta^{\mu\nu}$,
so that $\gamma^0$ is Hermitian and $\gamma^i$ are anti-Hermitian. Also, $\gamma_5 \equiv i \gamma^{0}\gamma^{1}\gamma^{2}\gamma^{3}$ is Hermitian.}
\be
S[\psi, \phi,\phi_5,A,A_5]
=
\int d^4x\,
\bar\psi
\Bigl[
i \slpartial
- m 
-g\phi
-i g_5\gamma_5\phi_5
-e\slash A
-e_5\gamma_5
{\slash A}_5
\Bigr]
\psi
\, .
\label{Sspinhalfmink}
\ee\no
In the following we absorb the coupling constants into the
background fields. 

Although those representations have been successfully tested on the calculation of amplitudes \cite{haasch} as well as of the Standard Model effective action \cite{hekosc1,hekosc2},
they are somewhat less appealing than their non-chiral QED and QCD analogues in so far as they imply a separate treatment of parity-even and
parity-odd amplitudes. For the euclidean\footnote{
In euclidean space we use anti-Hermitian gamma matrices with 
$\lbrace \gamma_E^\mu,\gamma_E^\nu \rbrace = - 2\delta^{\mu\nu}$
(note that this differs from \cite{29,33,41} where $\lbrace \gamma_E^\mu,\gamma_E^\nu \rbrace = +2\delta^{\mu\nu}$
was used) and Hermitian $\gamma_{5E} = \gamma_E^1\gamma_E^2\gamma_E^3\gamma_E^4$.
The subscript ``$E$'' is omitted in the following.
The euclidean $\varepsilon$-tensor is defined by
$\varepsilon_{1234} = 1$.
} effective action
corresponding to integrating out the fermion degrees of freedom in \eqref{Sspinhalfmink},  
\be
\Gamma_{\rm E}[\phi,\phi_5,A,A_5] = \ln {\rm Det}
\Bigl[{\slash p}_E +m+ \phi + i\gamma_5\phi_5 +{\slash A}_E
+\gamma_5{\slash A}_{E5}\Bigr] \,,
\label{defEAeuc-annchiral}
\ee\no
this shows up in a separate treatment of its real and imaginary parts
(the latter in this context arises only for parity-odd amplitudes
as an artefact of the Wick rotation of the Levi-Civita tensor). 

For the special case of only the vector and axialvector couplings, 
it was shown by two of the present authors in \cite{29} 
how to avoid the separation of the effective action into its real and imaginary parts
by formally rewriting it in terms of the effective action for a scalar loop in a fictitious
non-abelian background. This made it possible to use known results for the expansion of the non-abelian heat-kernel for analysing the anomalous and non-anomalous content of the heat-kernel expansion for the vector-axialvector case.  More significantly, it allowed the reuse of the worldline representation of the non-abelian effective action 
to derive a novel such representation for the effective action in the vector-axialvector background. 

This representation was then used in \cite{33} for deriving a Master Formula for one-loop vector-axialvector amplitudes,
as well as for a recalculation of the ABJ anomaly \cite{adler,beljac}, with the interesting result that in this particular 
worldline formulation the anomalous divergence naturally appears at the axialvector current, not the vector one. 
It is straightforward to incorporate an additional constant electromagnetic background field along the lines of 
\cite{shaisultanov,18}, which was used for a calculation of the one-loop vector-axialvector polarisation tensor \cite{42}
and the axialvector-axialvector polarisation tensor \cite{47} in such a field. See also \cite{Mansfield, Copinger} for worldline models of chiral fermions in the Standard Model and \cite{WLUnif} for a generalisation to unified theories. 

In the present Letter, we generalise the approach of \cite{29} to the closed fermion loop coupled to the full set of 
scalar, pseudoscalar, abelian vector and axialvector fields. This turns out to be a by no means straightforward task since, differently from the pure vector-axialvector
case, the resulting worldline action contains terms with an odd number of Grassmann fields, which require the use of a suitable path-ordering.
Moreover, although the model is defined in flat space, at the operatorial level the axial fields enter coupled to products of gamma matrices which, when passing to the path integral, imply an ordering ambiguity which must be resolved. 
Following strategies developed in the curved-space context \cite{Bastianelli:2006rx},
we employ the Time Slicing regularisation of the path integral which requires the addition of a  specific set of counterterms to the worldline Lagrangian -- which we calculate completely and \textit{non-perturbatively} -- and provides 
unambiguous rules for the treatment of products of distributions occurring in certain diagrams of the one-dimensional worldline theory. 
We compute various scattering amplitudes and heat-kernel coefficients, finding perfect agreement with space-time QFT approaches and heat-kernel techniques respectively.

The outline of the manuscript is as follows: we start by deriving a proper-time representation of the effective action in Section \ref{secWLGamma}. We then 
use standard coherent-state methods 
to obtain a worldline description in Section \ref{sec:Worldline representation of the effective action}, where we also 
calculate the counterterms corresponding to our path integral regularisation. In Section \ref{secHKExp} we employ the worldline representation to rederive
some heat kernel coefficients as a basic test of our formalism.
In Section \ref{sec-amps} we lay out the rules for the calculation of one-loop amplitudes with an arbitrary number of scalar, pseudoscalar, vector and axialvector
legs in the formalism. Those we apply in Section \ref{sec-even} to some parity-even, in Section \ref{sec-odd} to some parity-odd amplitudes.
Section \ref{sec-conc} offers a summary and outline of possible further extensions. 
In the appendix we use the formalism of Section 
 \ref{sec:Worldline representation of the effective action},
for calculating the heat-kernel coefficients $a_1$ and $a_2$ in a general (even) dimension.
We also briefly comment on the consequences of chiral invariance on the amplitudes and heat-kernel coefficients in this model.

\section{Proper-time representation of the effective action}
\label{secWLGamma}

The euclidean effective action corresponding to the action \eqref{Sspinhalfmink} can be expressed as the functional determinant 
\be
\Gamma[\phi,\phi_5,A,A_5] = \ln {\rm Det}
\bigl\lbrack m+  {\cal O} \bigr\rbrack
\label{defEAeuc-annchiral-1}
\ee\no
with the euclidean Dirac operator
\bear
{\cal O} = 
{\slash p} +{\slash A}+\gamma_5{\slash A}_{5}  + \phi + i \gamma_5\phi_5  \,.
\label{calO}
\ear
Note that, in general, this operator is neither Hermitian nor anti-Hermitian. 
Using $\gamma_5^2=1$ and the cyclic invariance of the determinant, we can rewrite this as
\bear
{\rm Det}\bigl(m+{\cal O} \bigr) = 
{\rm Det}\bigl\lbrack \gamma_5^2 \bigl(m+ {\cal O} \bigr) \bigr\rbrack = 
{\rm Det}\bigl\lbrack \gamma_5 \bigl(m+{\cal O}\bigr) \gamma_5 \bigr\rbrack = 
{\rm Det}\bigl(m+\tilde {\cal O} \bigr)\,,
\nonumber\\
\label{14-cyclic}
\ear
where
\bear
\tilde {\cal O} := \gamma_{5} \mathcal{O} \gamma_{5} = 
-{\slash p} -{\slash A}-\gamma_5{\slash A}_{5}  + \phi + i \gamma_5\phi_5 \,.
\label{14-defcalOtilde}
\ear
Thus without loss of generality we have the alternative representation
\bear
\Gamma[\phi,\phi_5,A,A_5] 
&=& \half \ln {\rm Det}\bigl\lbrack (m + {\cal O})(m+\tilde{\cal O}) \bigr\rbrack \,.
\label{14-Gammafin}
\ear
A long but straightforward calculation shows that we can further write this in the second order form
\be
(m + {\cal O})(m+\tilde{\cal O}) 
=
m^2 
- (\partial_\mu + i{\mathfrak A}_\mu)^2 + {\mathfrak a}\,,
\label{14-newKG}
\ee\no
where the matrix-valued potentials take the form ($\gamma^{\mu\nu}:=\frac12[\gamma^\mu,\gamma^\nu]$)
\bear
{\mathfrak A}_\mu &=&  A_{\mu} +\gamma_{\mu\nu} 
\gamma_5
A_5^{\nu}
   + i \gamma_\mu \gamma_5 \phi_5 \, ,
   \nonumber\\
{\mathfrak a} &=& 
  \frac{i}{2} \gamma^{\mu\nu} 
\bigl(
\partial_{\mu}A_{\nu}-\partial_{\nu}A_{\mu}
\bigr)
+i\gamma_5 \partial_{\mu}A_{5}^{\mu} + (D-2) A_5^2 
+ (D-1)\phi_5^2 
\nonumber\\
&&
+2i(D-2)\phi_5 {\slash A}_5 
-i\gamma^\mu \partial_\mu \phi +\phi^2 
+2i\phi \phi_5 \gamma_5  +2m (\phi+i\gamma_5\phi_5)
\, .
\nonumber\\
\label{14-calAW}
\ear
The operator on the RHS of (\ref{14-newKG}) resembles a Klein-Gordon Hamiltonian with matrix-valued potentials (as is already familiar from the worldline formalism applied to QED -- see \cite{Morgan:1995te, Guzman:2023hzq} for a discussion of the transition from the first- to the second-order formalism of the Dirac theory).  Using \eqref{14-Gammafin}, \eqref{14-newKG} together with the ``Tr ln = ln Det'' identity and the integral representation of the logarithm, we 
obtain the proper-time representation of the effective action in the form
\be
\Gamma[\phi,\phi_5,A,A_5] = -\half\, \Tr\, \int_0^\infty \, 
\frac{dT}{T} \, \exp 
\Bigl\lbrace
- T \left[-
(\partial_\mu + i{\mathfrak A}_\mu)^2 + {\mathfrak a} + m^2 \right]
\Bigr\rbrace\, .\\
\label{14-EAHK}
\ee
Note that the exponent is not Hermitian, which is the price to pay for writing down the whole effective action 
-- which in general is not real -- in one piece. 

As it stands, the representation \eqref{14-EAHK} is already quite useful as a tool for obtaining 
the Seeley-DeWitt (or heat-kernel) expansion
for the case at hand, which reads
\cite{dewitt,bles,vandeven,vandeven-indexfree,25,vassilevich}:

\be
 \Tr \exp 
\Bigl\lbrace
- T \left[-
(\partial_\mu + i{\mathfrak A}_\mu)^2 + {\mathfrak a}  \right]
\Bigr\rbrace
 = \frac{1}{(4\pi T)^\frac{D}{2} }\, \sum_{n=0}^{\infty} T^n
\int d^Dx \, 
a_n 
(x,x)\,.
\label{DeWitt}
\ee\no
In the Appendix, we show how \eqref{14-EAHK} is well adapted for the determination of the heat-kernel coefficients, $a_{n}$, and we provide the first $3$ coefficients explicitly in arbitrary dimension. We will also use these explicit results in the following section as a means to verify the calculation of some counterterms required in the time-slicing regularisation of worldline representation of the effective action.

\section{Worldline representation of the effective action}
\label{sec:Worldline representation of the effective action}

Although the representation \eqref{14-EAHK} is already quite useful for a ``standard'' computation of the effective action,
our real aim is to find a worldline representation, which as usual should be universally applicable to the computation of the
effective action itself, as well as of the associated amplitudes. 

\subsection{Naive worldline Lagrangian}

By applying the coherent-state formalism \cite{ohnkas,hehoto} along the lines of \cite{29,33,41}, \eqref{14-EAHK} transforms 
into the following worldline representation of the effective action:
\bear
\Gamma [\phi,\phi_5,A,A_5]
= -\half\, \int_0^\infty \, 
\frac{dT}{T} 
\e^{-m^2T}
\int_P Dx
\int_{A/P} D\psi \, 
{\cal P}
\e^
{
-\int_0^Td\tau\,
L(\tau)
}\,,
\nonumber\\
\label{14-Gammawl}
\ear
with worldline Lagrangian
\bear
L(\tau)
&=&
\fourth \dot x^2
+\half\psi_{\mu}\dot\psi^{\mu}
+i\dot x^{\mu}A_{\mu}
-i\psi^{\mu}F_{\mu\nu}\psi^{\nu}
-2i\dot x^{\mu}\psi_{\mu}\psi_{\nu}A_5^{\nu}\hat\gamma_5
+i\partial_{\mu}A^{\mu}_5 \hat\gamma_5
\nonumber\\ 
&&+(D-2)A_5^2
+ \phi^2 + (D-1)\phi_5^2
+2i\phi\phi_5\hat\gamma_5
+ 2m(\phi+i\phi_5\hat\gamma_5) 
\nonumber\\
&&-\sqrt{2} \psi^{\mu}\partial_{\mu}\phi
+\sqrt{2}i \dot x_{\mu}\psi^{\mu} \phi_5\hat\gamma_5
+2\sqrt{2}(D-2)\psi_{\mu}A_5^{\mu} \phi_5
 \, .
\label{14-Lnew}
\ear\no
Here as usual $\int_{P} {\cal D}x$ 
denotes the coordinate path integral over the space
of all closed (periodic) loops with fixed proper-time length $T$, and
$\int_{A/P}{\cal D}\psi$ 
a path integral over Grassmann-valued functions, which are the eigenvalues of operators identified with the $\gamma$ matrices through $\hat{\psi}^{\mu} =\frac{i}{\sqrt{2}}\gamma^{\mu}$.
The periodicity properties of 
$\int{\cal D}\psi$ 
are perturbatively determined by the number of interactions with the axialvector field $A_5$; 
antiperiodic (periodic) boundary conditions
on $\psi$, $\psi(T) = - (+)\,\psi(0)$, have to be used
if that number is even (odd). 
This book-keeping is provided by the $\gamma_5$-matrix which has now turned into an operator $\hat\gamma_5$ 
that is doing nothing but switching the boundary conditions when the exponential in (\ref{14-Gammawl}) is expanded.

The worldline Lagrangian \eqref{14-Lnew} has an unusual feature which does not appear in the standard QED case, and not even in the pure vector--axialvector case
treated in \cite{29,33}, namely the appearance of terms linear in the Grassmann variables $\psi^{\mu}$. A careful analysis shows that this does not lead to
inconsistencies in the path-integral formalism, just to the following two complications: first, path-ordering must be used
whenever such terms are present in a term; second, 
translation invariance cannot be used in the usual way to arbitrarily fix the location of one of the vertices on the loop. 
Moreover, the operator $\hat\gamma_5$ in such cases assumes a slightly less trivial role since, whenever there is more than one such operator around,
they must be cancelled in pairs by bringing them next to each other, leading to relative signs whenever this requires
a $\hat\gamma_5$ to move across a single $\psi^\mu$. With these instructions, the worldline representation, (\ref{14-Gammawl}), and the operators appearing in the action (\ref{14-Lnew}) are well-defined, at least perturbatively in the fields. 

Finally, it will be useful to observe that, introducing the new field variables
\bear
\tilde A^\mu &\equiv & A^\mu + \sqrt{2} \psi^\mu \phi_5\hat\gamma_5 \\
\tilde A_5^\mu &\equiv & A_5^\mu + \sqrt{2} \psi^\mu \phi_5 \\
\tilde \phi &\equiv & \phi +i \phi_5\hat\gamma_5\,,
\ear 
the worldline Lagrangian \eqref{14-Lnew} can be rewritten more compactly as
\bear
L(\tau)
&=&
\fourth \dot x^2
+\half\psi_{\mu}\dot\psi^{\mu}
+i\dot x^{\mu}\tilde A_{\mu}
-i\psi^{\mu}\tilde F_{\mu\nu}\psi^{\nu}
-2i\dot x^{\mu}\psi_{\mu}\psi_{\nu}\tilde A_5^{\nu}\hat\gamma_5
+i\partial_{\mu}\tilde A^{\mu}_5\hat\gamma_5
-\sqrt{2} \psi^{\mu}\partial_{\mu}\tilde \phi 
\nonumber\\ &&
+2m \tilde\phi
+ \tilde\phi^2 
+ D\phi_5^2
+ (D-2) \tilde A_5^2
\, .
\label{14-Lcompact}
\ear\no
However, one technical aspect of the construction of the path integral remains unspecified: specifically how it should be defined as the continuum limit of some regularised object. In the following subsection we interpret the path integral in Time Slicing regularisation, and calculate the necessary counter terms which accompany this prescription.

\subsection{Counterterms from time-slicing regularisation}

It is well known that worldline path integrals in curved spaces need to be regularised, as the actions they involve are one-dimensional non-linear sigma models, and their short-time perturbation theory leads to ambiguities and divergences, which need to be treated. In the past, 
three main schemes have been developed, namely Time Slicing, Mode Regularisation and 
Worldline Dimensional Regularisation \cite{Bastianelli:2006rx}. 

In the present manuscript, the target space is flat and gravity is absent, but the vector potential which occurs in the point particle Hamiltonian is matrix-valued. In particular, the axial vector field enters coupled to the Lorentz generators in the spin $1/2$ representation, $\gamma_{\mu\nu}$, leading to an ``artificial'' spin connection term. Thus, some care is needed in order to pass from the operatorial approach to the corresponding path integral: divergences are not expected on the compact interval, but there are worldline integrals which turn out to be ambiguous, since they involve products of distributions. Moreover, as already pointed out, a path ordering is needed when Grassmann odd terms are present. In such a case it appears that the only viable regularisation scheme is Time Slicing as it is first-principled, and applicable at the level of the Hamiltonian operator. 

Time Slicing with the mid-point rule requires the Hamiltonian operator to be Weyl-ordered. In our case, in order to use the coherent-state approach to the fermionic operators this implies that
\begin{align}
\widetilde{\Pi}^{2} = (-i\partial_\mu + {\mathfrak A}_\mu)^2
\end{align}           
must be Weyl-ordered. The orbital part is naturally Weyl-ordered as the space is flat. Thus, to achieve Weyl-ordering, only the following squared operator must be considered
\begin{align}
(\gamma_{\mu\nu} 
\hat\gamma_5 A_5^{\nu}   + i \gamma_\mu \hat\gamma_5 \phi_5)^2~,
\label{eq:squared}
\end{align}
containing both pseudoscalar and pseudovector fields (here, we keep using $\hat\gamma_5$ as a book-keeping device). 

Now, a given operator $\hat {\cal O}$ is Weyl-ordered when it is written in its (anti)-symmetrised form in its ladder operators plus a remainder, which is unambiguous in the product of ladder operators, i.e. $\hat {\cal O}=\hat {\cal O}_W=\hat {\cal O}_S+\Delta$, where $S$ stands for (anti)-symmetrised\footnote{Note that we use the notation of the book~\cite{Bastianelli:2006rx}, but also use some results adapted from~\cite{deBoer:1995cb}, where Weyl ordering is defined differently.}. Thus, in order to write the symmetrised version of a generic operator involving $\gamma$ matrices we have to write their corresponding Majorana operators $\hat{\psi}^{\mu}$  in terms of ladder operators, which satisfy the fermionic anticommutation algebra. In a generic even dimension, $D=2d$, this can be done as follows (we omit hats on the ladder operators to avoid cluttering)
\begin{align}
\chi^m:=\frac{1}{\sqrt2}\big( \hat \psi^m+i \hat\psi^{m+d}\big)\,,\quad \bar \chi^m:=\frac{1}{\sqrt2}\big( \hat \psi^m-i \hat\psi^{m+d}\big)\,,\quad m=1,\dots,d~.
\end{align} 
 In the present case we are obviously interested in the (anti)-symmetrisation of the Hamilton operator, which in the Time Slicing of the evolution operator, is linked to the mid-point rule for the corresponding Hamiltonian function -- see e.g. Ref.~~\cite{Bastianelli:2006rx}. With the above	 identification $\gamma_\mu=-i\sqrt2 \hat\psi_\mu$ which implies $\{\hat\psi_\mu,\hat\psi_\nu\} =\delta_{\mu\nu}$. In particular, anti-symmetrisation for a pair of these ladder operators takes the form $(\chi^{m}  \bar{\chi}^{n} )_{S} \equiv \frac{1}{2} \big( \chi^{m}  \bar{\chi}^{n} - \bar{\chi}^{n}  \chi^{m} \big)$.

Let us thus consider the different pieces arising from (\ref{eq:squared}) separately. 
We start with the Weyl-ordering of the square of the second addendum in~\eqref{eq:squared}, being
\begin{align}
(i \gamma_\mu \hat\gamma_5 \phi_5)_W^2 = -2\phi_{5}^{2}(\hat\psi_{\mu}  \hat\psi^{\mu} )_W = -2\phi_{5}^{2}(\hat\psi_{\mu}  \hat\psi^{\mu} )_S -D\phi_5^2~,
\end{align}  
where, of course, $(\psi^{\mu}\psi^{\nu})_{S} = \frac{1}{2}(\psi^{\mu}\psi^{\nu} - \psi^{\nu}\psi^{\mu})$,  i.e.
\begin{align}
(\hat\psi_\mu\hat\psi_\nu)_W=(\hat\psi_\mu\hat\psi_\nu)_S+\frac12 \delta_{\mu\nu}~,
\label{eq:rule}
\end{align}
which holds for a pair of Majorana fermionic operators (see also Eq. (55) of \cite{deBoer:1995cb}). Thus, from the expression above we have the following contribution to the counterterm,
\begin{align}
V_{TS}\supseteq -D\phi_5^2~,
\end{align}
where `TS' stands for Time Slicing. 

Let us now consider the Weyl-ordering of the cross-product term,
\begin{align}
i\phi_{5}A_{5}^{\nu}(-\gamma_{\mu\nu}\gamma^\mu+\gamma^\mu\gamma_{\mu\nu})_W = -2\sqrt2\phi_{5}A_{5}^{\nu} (-\hat\psi_{\mu\nu}\hat\psi^\mu+\hat\psi^\mu\hat\psi_{\mu\nu})_W~,
\end{align}
where $\hat{\psi}_{\mu\nu} := \frac{1}{2}[\hat{\psi}_{\mu}, \hat{\psi}_{\nu}]$. This involves the product of three operators, which was not worked out explicitly in Ref.~\cite{deBoer:1995cb}, so we outline some details of the computation. Thus,
\begin{align}
&2 (-\hat\psi_{\mu\nu}\hat\psi^\mu+\hat\psi^\mu\hat\psi_{\mu\nu})_W = (-2\hat\psi_\mu \hat\psi_\nu\hat\psi^\mu +\hat\psi_\nu\hat\psi_\mu \hat\psi^\mu+\hat\psi_\mu \hat\psi^\mu\hat\psi_\nu)_W \nonumber\\&=  (-2\hat\psi_\mu \hat\psi_\nu\hat\psi^\mu +\hat\psi_\nu\hat\psi_\mu \hat\psi^\mu+\hat\psi_\mu \hat\psi^\mu\hat\psi_\nu)_S+\Delta~.
\end{align}
Let us, for instance, consider the first term here, with $\nu=n$ (the case $\nu=n+d$ can be obtained identically),
\begin{align}
(\hat\psi_\mu \hat\psi_n\hat\psi^\mu)_S=(\hat\psi_m \hat\psi_n\hat\psi^m)_S+(\hat\psi_{m+d}\, \hat\psi_n\, \hat\psi^{m+d})_S~,
\end{align}
which, in terms of the ladder operators reduces to,
\begin{align}
(\hat\psi_\mu \hat\psi_n\hat\psi^\mu)_S&=\frac1{2\sqrt2} \Big[\big(\chi_m+\bar\chi_m \big)\big(\chi_n+\bar\chi_n \big)\big(\chi^m+\bar\chi^m \big)\Big]_S\nonumber\\
&-\frac1{2\sqrt2} \Big[\big(\chi_m-\bar\chi_m \big)\big(\chi_n+\bar\chi_n \big)\big(\chi^m-\bar\chi^m \big)\Big]_S\nonumber\\
&=\frac1{\sqrt2}\big(  \chi_m \chi_n\bar\chi^m +\chi_m\bar\chi_n\bar\chi^m + \bar\chi_m\chi_n\chi^m+\bar\chi_m\bar\chi_n\chi^m\big)_S~.
\end{align}
Explicitly, the first term can be written as
\begin{align}
\big(  \chi_m \chi_n\bar\chi^m\big)_S =\frac{1}{3!}\big(2\chi_m \chi_n\bar\chi^m +2\bar\chi^m\chi_m \chi_n-\chi_m\bar\chi^m \chi_n+\chi_n\bar\chi^m \chi_m\big)_S 
=\chi_m \chi_n\bar\chi^m +\frac{d-1}{2}\chi_n~,
\end{align}
by using the anticommutation relations among the ladder operators. Thus, the latter
can be rewritten as
\begin{align}
\big(\chi_m \chi_n\bar\chi^m\big)_W=\big(  \chi_m \chi_n\bar\chi^m\big)_S-\frac{d-1}{2}\chi_n~.
\end{align}
One proceeds with similar steps with the other three terms and the final result is cast in terms of the original $\psi$s as
\begin{align}
(\hat\psi_\mu \hat\psi_n\hat\psi^\mu)_W=(\hat\psi_\mu \hat\psi_n\hat\psi^\mu)_S-\frac{D-2}{2}\hat\psi_n~.
\end{align}
Obviously, the same rule holds if we replace $n$ with $n+d$, i.e.
\begin{align}
(\hat\psi_\mu \hat\psi_\nu\hat\psi^\mu)_W=(\hat\psi_\mu \hat\psi_\nu\hat\psi^\mu)_S-\frac{D-2}{2}\hat\psi_\nu~.
\end{align}
Proceeding in the exact same way, one also obtains
\begin{align}
& (\hat\psi_\mu\hat\psi^\mu \hat\psi_\nu)_W=(\hat\psi_\mu \hat\psi^\mu \hat\psi_\nu)_S+\frac{D}{2}\hat\psi_\nu~,\\
& (\hat\psi_\nu \hat\psi_\mu\hat\psi^\mu )_W=(\hat\psi_\nu\hat\psi_\mu \hat\psi^\mu )_S+\frac{D}{2}\hat\psi_\nu~,
\end{align}
and the final result reads,
\begin{align}
-2\sqrt2 (-\hat\psi_{\mu\nu}\hat\psi^\mu+\hat\psi^\mu\hat\psi_{\mu\nu})_W =-2\sqrt2 (-\hat\psi_{\mu\nu}\hat\psi^\mu+\hat\psi^\mu\hat\psi_{\mu\nu})_S-2\sqrt2 (D-1)\hat\psi^\nu~. 
\end{align}
Thus, the cross term in question requires the counterterm contribution
\begin{align}
V_{TS}\supseteq -2\sqrt2 (D-1)\psi\cdot A_5\phi_5~.
\end{align}
Finally, let us consider the square of the first addendum, for which we have
\begin{align}
 (\gamma_{\mu\nu} 
\hat\gamma_5 A_5^{\nu})^2_W &=
(\gamma_{\mu\nu}  \gamma_{\mu\nu'} )_W A_5^{\nu}A_5^{\nu'}\nonumber\\
&= 4(\hat\psi_{\mu\nu}  \hat\psi_{\mu\nu'} )_W A_5^{\nu}A_5^{\nu'}~.
\end{align}
By defining the ``artificial'' spin connection,
\begin{align}
\omega^l{}_{\mu\nu} =A_{5\, \mu}\delta^l_\nu - A_{5\, \nu}\delta^l_\mu\,, 
\end{align}
 the expression above can be re-written as
\begin{align}
\omega^l{}_{\mu\nu}\, \omega^l{}_{\mu'\nu'} (\hat\psi^{\mu\nu}  \hat\psi^{\mu'\nu'} )_W~.
\end{align}
At this point we can directly use the results obtained in \cite{deBoer:1995cb}, cf. Eq. (56), for a spinning particle in curved space, i.e.
\begin{align}
\omega^l{}_{\mu\nu}\, \omega^l{}_{\mu'\nu'} (\hat\psi^{\mu\nu}  \hat\psi^{\mu'\nu'} )_W&=\omega^l{}_{\mu\nu}\, \omega^l{}_{\mu'\nu'} (\hat\psi^{\mu\nu}  \hat\psi^{\mu'\nu'} )_S + \frac12 \omega^l{}_{\mu\nu}\omega^l{}^{\mu\nu}\nonumber\\
& =\omega^l{}_{\mu\nu}\, \omega^l{}_{\mu'\nu'}(\hat\psi^{\mu\nu}  \hat\psi^{\mu'\nu'} )_S-(D-1)A_5^2~.
\end{align}
The final term on the RHS will thus sum to the counterterm potential, which globally now reads
\begin{align}
V_{TS} =-D\phi_5^2-2\sqrt 2 (D-1)\psi_\mu A_5^\mu \phi_5 -(D-1)A_5^2~.
\end{align}
Due to the superrenormalisability of the worldline action this is already the full non-perturbative counterterm Lagrangian. Indeed, the theory is not divergent on the compact interval as there are no derivative interactions. There are, however, ambiguities as we will illustrate below, but these are fully resolved by the Time Slicing prescription and the counterterm potential. We have checked that no higher order contributions to the counterterm potential can arise by an analysis similar to the one performed in \cite{Bastianelli:2006rx} for the curved-space path integral (in brief, the only additional fields that enter higher loop worldline diagrams are $x(\tau)$s, without derivatives, which improve their behaviour).

\subsection{Total worldline Lagrangian}

We must include the TS counterterm potential, calculated above, in the worldline action. Hence, there is a shift
\begin{align}
L(\tau) \rightarrow L(\tau) + V_{TS} &= \fourth \dot x^2
+\half\psi_{\mu}\dot\psi^{\mu}
+i\dot x^{\mu}A_{\mu}
-i\psi^{\mu}F_{\mu\nu}\psi^{\nu}
-2i\dot x^{\mu}\psi_{\mu}\psi_{\nu}A_5^{\nu}\hat\gamma_5
+i\partial_{\mu}A^{\mu}_5 \hat\gamma_5
\nonumber\\ 
&-A_5^2
+ \phi^2  -\phi_5^2
+2i\phi\phi_5\hat\gamma_5
+ 2m(\phi+i\phi_5\hat\gamma_5) 
\nonumber\\
&
-\sqrt{2} \psi^{\mu}\partial_{\mu}\phi
+\sqrt{2}i \dot x_{\mu}\psi^{\mu} \phi_5\hat\gamma_5
-2\sqrt{2}\psi_{\mu}A_5^{\mu} \phi_5
 \, .
 \label{eqLCT}
\end{align}
Remarkably, the time slicing counterterms have the effect of removing all explicit $D$-dependence from the worldline action!
 With respect to the shifted field variables we get
\begin{align}
L(\tau) + V_{TS}
&=
\fourth \dot x^2
+\half\psi_{\mu}\dot\psi^{\mu}
+i\dot x^{\mu}\tilde A_{\mu}
-i\psi^{\mu}\tilde F_{\mu\nu}\psi^{\nu}
-2i\dot x^{\mu}\psi_{\mu}\psi_{\nu}\tilde A_5^{\nu}\hat\gamma_5
+i\partial_{\mu}\tilde A^{\mu}_5\hat\gamma_5
\nonumber\\ 
&-\sqrt{2} \psi^{\mu}\partial_{\mu}\tilde \phi  +2m \tilde\phi
+ \tilde\phi^2 
- \tilde A_5^2
\, .
\label{LCTShift}
\end{align}
Notice there is now no longer a $\tilde\phi_{5}^{2}$ contribution.

\section{Worldline calculation of the heat-kernel expansion} 
\label{secHKExp}

\subsection{Worldline path-integration rules}

With the worldline Lagrangian in hand, we can now apply the usual worldline machinery \cite{5,41} to calculate the
heat-kernel expansion of the effective action. First, the zero mode of the coordinate path integral is fixed 
by splitting $x^\mu (\tau) = x_0^\mu +  q^\mu (\tau)$ with the 
loop average position $x_0^\mu \equiv \frac{1}{T} \int_0^Td\tau\,  x^\mu (\tau)$. Then all external fields are Taylor-expanded about $x_0$,
resulting in Gaussian path integrals that can be evaluated by Wick contractions using the elementary correlator 
\bear
\langle q^\mu(\tau)q^\nu(\tau') \rangle = - G(\tau,\tau') \delta^\mn \, ,
\label{9-wickG}
\ear
with the ``string-inspired'' Green's function adapted to the space orthogonal to the zero mode
\bear
\quad G(\tau,\tau') = \vert \tau - \tau' \vert - \frac{(\tau-\tau')^2}{T} \, .
\label{defG}
\ear
As we already mentioned above, the evaluation of the Grassmann path integral depends on whether 
a given insertion under the path integral is of even or odd parity, that is, whether it has an even or odd number of $\hat\gamma_5$ factors.
In the even case the Grassmann path integral is computed with anti-periodic boundary conditions 
like in the familiar QED case, using the worldline correlator
\bear
\langle \psi^{\mu}(\tau_1)\psi^{\nu}(\tau_2)\rangle &=& \half G_F(\tau,\tau')\delta^{\mu\nu} \, , \quad G_F(\tau,\tau') = {\rm sgn}(\tau-\tau')\,,
\label{11-wickgrass2point}
\ear
and the free path integral over the $\psi$'s gives $2^{\frac D2}$, since it is nothing but the trace of the identity in the fermionic Hilbert space, i.e. the number of spinorial components of a Dirac field.

In the parity-odd case, however, the boundary conditions on the Grassmann path integral become periodic, 
so that one now encounters a fermionic zero mode, which must be removed before executing the path integral
by factorising the Hilbert space of periodic Grassmann functions into the constant functions $\psi_0$ and their orthogonal complement
$\xi(\tau)$,
\bear
\int_P{\cal D}\psi &=& \int d\psi_0\int {\cal D}\xi \, , \\
\psi^{\mu}(\tau)&=&\psi_0^{\mu} + \xi^{\mu}(\tau)\, ,\label{splitpsi}\\
\frac{1}{T}\int_0^Td\tau\,\xi(\tau) &=& 0 \, .
\label{splitgrass}
\ear\no
The zero mode integration then produces a Levi-Civita tensor,
\be
\int d^4\psi_0\, 
\psi_0^{\mu}\psi^{\nu}_0\psi^{\kappa}_0\psi^{\lambda}_0 
=\varepsilon^{\mu\nu\kappa\lambda}\, ,
\label{zeromodeintegral}
\ee\no
(respectively its $D$-dimensional generalisation)
whose appearance is expected in the parity-odd case. 
The path integral over $\xi$ has to be performed using the correlator
\be
\langle\xi^{\mu}(\tau)\, \xi^{\nu}(\tau')\rangle
=
\delta^\mn \bigl\la \tau | \Bigl(\frac{d}{d\tau}\Bigr)^{-1} | \tau' \bigr\ra_{SI}
=\delta^{\mu\nu}\half\dot G (\tau,\tau')
\label{xicorr}
\ee\no
and the free path integral over $\xi$ is normalised to unity.

It is important to emphasise that calculating the expansion \eqref{14-EAHK} of the diagonal elements of the heat-kernel in this way 
will, in general, produce expansion coefficients that differ from the ``standard''  heat-kernel ones by certain total-derivative terms
\cite{25}. This is an effect of the ``string-inspired'' Green's function \eqref{defG} and could be avoided by instead using
the one corresponding to Dirichlet boundary conditions \cite{mckeon},
\bear
\Delta(\tau,\tau') &=&
\frac{\vert\tau - \tau' \vert}{2} - \frac{\tau + \tau'}{2} +  \frac{\tau\tau'}{T}  \;.
\label{9-defDeltaT}
\ear
However, higher-order calculations in scalar and gauge theories have shown \cite{6,25} that the use of the string-inspired Green's function 
is generally preferable since it avoids the breaking of Bose symmetry which makes it easier to reduce the result to a minimal basis of invariants. 

\subsection{Heat kernel expansion}

Consulting the heat kernel coefficients provided in the Appendix, we have verified that this procedure correctly reproduces all the terms given in \eqref{computea12}. However, let us show some examples that illustrate some of the novel aspects of our approach here. We fix $m = 0$ in this section for consistency with the calculations presented in the Appendix.

Firstly, we demonstrate one of the unusual terms linear in Grassmann variables: the first term on the last line of the worldline action (\ref{eqLCT}), that at quadratic order contributes to $2^{\frac{D}{2}-1}(\partial \phi)^{2} \subseteq a_{2}$. Expanding the exponential in (\ref{14-Gammawl}), the contributing term is\footnote{Note that there is no contribution to $a_{2}$ from expanding the $\phi^{2}(x(\tau))$ term in the action to first order about $x_{0}$, since this provides an orbital contraction $\langle q^{\mu}(\tau)q^{\nu}(\tau)\rangle = -\eta^{\mu\nu}G(\tau, \tau) = 0$.}
\begin{equation}
	2^{\frac{D}{2}} \Big\langle \mathcal{P}\Big(\int_{0}^{T}d\tau\, \psi(\tau)\cdot \partial \phi(x_{0}) \Big)^{2} \Big\rangle = 2^{\frac{D}{2}+1} \int_{0}^{T}d\tau \int_{0}^{\tau}d\tau' \big \langle \psi^{\mu}(\tau) \psi^{\nu}(\tau') \big \rangle \partial_{\mu}\phi \partial_{\nu} \phi \, .
\end{equation}
With anti-periodic boundary conditions the contraction of fields gives $\frac{1}{2}\delta^{\mu\nu}\sigma(\tau - \tau') \rightarrow \frac{1}{2}\delta^{\mu\nu}$ and so the integral evaluates to $\frac{T^{2}}{4} \delta^{\mu\nu}$, reproducing the correct result for this heat kernel coefficient (without the path ordering, the coefficient would have vanished). The same works for the term term in $a_{2}$ proportional to $\phi^{2}\phi_{5}^{2}$. This time there are several contributions: 
\begin{itemize}
	\item From the cross-term from expanding the $\phi^{2}$ and $\phi_{5}^{2}$ terms in (\ref{eqLCT}) to first order: \\
	$-2^{\frac{D}{2}}\int_{0}^{T}d\tau \, \phi^{2} \int_{0}^{T}d\tau' \, \phi_{5}^{2} = -2^{\frac{D}{2}}T^{2} \phi^{2} \phi_{5}^{2}$.
	\item From expanding the $\phi \phi_{5}$ term in the action to second order: \\
	$-2^{\frac{D}{2}+1}\big(\int_{0}^{T}d\tau \, \phi \phi_{5}\big)^{2} = -2^{\frac{D}{2}+1}T^{2}\phi^{2}\phi_{5}^{2}$.
	\item From the cross term from expanding the $\phi^{2}$ term to first order and the $\phi_{5}\hat{\gamma}_{5}$ term to second order. Now the path ordering is needed again, and a sign is picked up from anti-commuting a $\hat{\gamma}$ past a $\psi$:\\
		$-2^{\frac{D}{2}+1}\int_{0}^{T}d\tau \, \phi^{2} \int_{0}^{T}d\tau' \int_{0}^{\tau'}d\tau'' \, \big \langle \dot{x}(\tau') \cdot \psi(\tau') \, \dot{x}(\tau'')\cdot \psi(\tau'')\big\rangle	 $.\\
		The contractions provide $\frac{D}{2}\ddot{G}(\tau, \tau')\sigma(\tau - \tau')$, which for reasons outlined below provides only a $-\frac{D}{T}$, so that overall this contribution evaluates to $2^{\frac{D}{2}}DT^{2} \phi^{2}\phi_{5}^{2}$.
\end{itemize}
Putting these contributions together correctly reproduces the coefficient $2^{\frac{D}{2}}T^{2}(D-3)\phi^{2}\phi_{5}^{2}$.

Already this calculation has thrown up a question of how to define the path integral computation properly. In order to check the validity of our construction it is now time to show how the Time Slicing regularisation and its associated counterterms resolve ambiguities that can arise (specifically with ill-defined products of distributions). So here we carry out further tests based on the heat-kernel coefficients given in the Appendix, focussing in particular on those which receive corrections from the worldline counterterms. It is sufficient to verify the axial terms present in the $a_1$ coefficient in the heat kernel expansion (i.e. $A_5^2$ and $\phi_5^2$), and the term $\phi_5 A_5^\mu \partial_\mu\phi$, which is part of the $a_2$ coefficient -- see \ref{secApp}.

Let us first consider the $\phi_5^2$ term, which gets a contribution $2^{\frac D2}$ from the quadratic term in the worldline action, and a contribution from the vertex linear in $\phi_5$, taken at the second order. The latter yields,
\begin{align}
-2^{\frac D2} (-\sqrt2 i)^2  D \phi_5^2  &\int_0^Td\tau \int_0^\tau d\tau' \ddot G(\tau-\tau')\frac12{\rm sgn}(\tau-\tau')~\\
=2^{\frac{D}{2}+1} T D\phi_{5}^{2} &\int_0^Td\tau \int_0^\tau d\tau'  \big(\delta(\tau - \tau') - \frac{1}{T}\big){\rm sgn}(\tau-\tau')
\label{eqPhi5Test}
\end{align}
(all fields are evaluated at $x_0$).
The previous integral would be ambiguous without a regularisation prescription, since it involves a product of distributions. In the TS  
prescription one treats the delta function and sign function as continuous limits of discretised expressions (see, for example, \cite{deBoer:1995cb}). In particular, in this case the delta term gives a vanishing contribution as it saturates the sign on the central point, and $\textrm{sgn}(0) = 0$ in Time Slicing. We thus get $2^{\frac D2}(-DT)\phi_5^2$, which combined with the $\phi_{5}^{2}$ term mentioned above provides (stripping off the factor of $T$ to extract the heat kernel coefficient)
\begin{align}
2^{\frac D2} (-(D-1))\phi_5^2\subseteq a_1~,
\end{align}  
which matches the heat-kernel computation reported in the Appendix. Note that without the $\phi_{5}^{2}$ counterterm the contribution (\ref{eqPhi5Test}) would have been multiplied by $-(D-1)$ and we would not have reproduced the expected result.

Secondly we consider the  $A_5^2$ term, which in the worldline approach gets the following contributions
\begin{align}
&2^{\frac D2}\Big\{ TA_5^2
+\frac1{2} \delta^{\mu\mu'}\int_0^Td\tau \int_0^Td\tau'\ddot G(\tau-\tau')(\delta_{\mu\mu'}\delta_{\nu\nu'}-\delta_{\nu\mu'}\delta_{\mu\nu'})G_F^2(\tau-\tau') A_5^\nu A_5^{\nu'}\Big\} \nonumber\\
=\,&2^{\frac D2}TA_5^2\Big\{ 1
+\frac{(D-1)}{T}\int_0^Td\tau \int_0^Td\tau'\Big(\delta(\tau-\tau')-\frac1T\Big) {\rm sgn}^2(\tau-\tau')\Big\}\,, 
\label{eqA5Test}
\end{align}
where again the delta function part yields a vanishing contribution. The leftover term is unambiguous and the final result thus reads,
\begin{align}
2^{\frac D2} (-(D-2))A_5^2\subseteq a_1 ~,
\label{ctA5}
\end{align}
which coincides with the heat kernel derivation -- again we strip off the $T$ factor to single out the contribution to the heat-kernel coefficient. Note for the second time that the $A_{5}^{2}$ counterterm is critical so that the contribution of the first term in (\ref{eqA5Test}) comes with the correct relative factor.

Finally let us consider a test for the third contribution to the counterterm. Obviously the latter does not contribute to $a_1$ as the v.e.v. of a single $\psi$ vanishes. Indeed there are no terms involving both $A^\mu_5$ and $\phi_5$ in $a_1$. The next simplest term to consider is $\phi_5 A_5\cdot\partial \phi$ which is part of $a_2$. This term comes from two contributions from the worldline path integral with action~\eqref{eqLCT}. The first one is the two-vertex contribution,
\begin{align}
2^{\frac D2}2!\ 4 \int_0^Td\tau \int_0^\tau d\tau'  \big\langle\psi^\mu(\tau)\psi^{\mu'}(\tau')\big\rangle \partial_\mu \phi\, A_{5\mu'} = 2^{\frac D2}\, 2\, \phi_5\, A_5\cdot \partial\phi\, T^2~.
\end{align}   
The second-one is the three-vertex contribution,
\begin{align}
&2^{\frac D2}\, 3!\, 4 \int_0^Td\tau \int_0^\tau d\tau'\int_0^{\tau'} d\tau{}''\big\langle \dot x^\mu \psi_\mu\psi_\nu(\tau)\, \psi^{\mu'}(\tau')\, \dot x^{\mu''}\psi_{\mu''}(\tau'')\big\rangle\, A_5^\nu\, \partial_{\mu'}\phi\, \phi_5\nonumber\\
& = 2^{\frac D2}\, 3!\,\int_0^Td\tau \int_0^\tau d\tau'\int_0^{\tau'} d\tau{}''\ddot G(\tau-\tau''){\rm sgn}(\tau-\tau'){\rm sgn}(\tau-\tau'')(D-1)\, \phi_5\, A_5\cdot \partial\phi\nonumber\\
&  =-2(D-2)\phi_5\, A_5\cdot \partial\phi\, T^2~,
\end{align}
where, once again, by TS rules, the delta function inside $\ddot G$ does not contribute. The integral over the simplex cancels the $3!$, and
 stripping off the $T^2$ factor as usual, we are thus left with \begin{align}
2^{\frac D2}(-2(D-1)) A_5\cdot \partial\phi \subseteq a_2~,
\end{align} 
 which again matches the heat kernel computation. 

\section{Amplitudes and vertex operators}
\label{sec-amps}

Somewhat more work is required to obtain the (one-particle irreducible) one-loop scattering amplitudes for external scalar, pseudoscalar, vector and axialvector particles from the effective action. To obtain the amplitudes in the worldline formalism, we must extract them from an appropriate linearisation of the effective action after specialising the background fields to plane waves representing the asymptotic states of the particles. 

\subsection{Linearisation and vertex operators}

From a systematic point of view, it will be useful to first linearise the various non-linear terms in the Lagrangian, being the last two terms in the version \eqref{LCTShift}.
Introducing auxiliary fields $z(\tau)$ and $Z^\mu (\tau)$, we can rewrite
\bear
\exp \Bigl[-\int_0^Td\tau\, \tilde \phi^2\Bigr]
&=&
\int D z (\tau)  \exp \Bigl[-\int_0^Td\tau 
\Bigl({z^2\over 4}+i z \tilde \phi\Bigr)
\Bigr]
\nonumber\\
\exp \Bigl[+\int_0^Td\tau \,\tilde A_5^2\Bigr]
&=&
\int D Z(\tau) \exp \Bigl[-\int_0^Td\tau 
\Bigl({Z^2\over 4}+\,Z\cdot \tilde A_5\Bigr)
\Bigr]\,.
\nonumber\\
\ear\no
The Wick contraction rules for these auxiliary fields are simply
\bear
\langle z(\tau)z(\tau')\rangle
&=&
2\delta (\tau-\tau')
\nonumber\\
\langle Z^\mu(\tau)Z^{\nu}(\tau')\rangle
&=&
2\delta^\mn \delta (\tau-\tau')\,,	
\nonumber\\
\label{14-wickaux}
\ear\no
and their free path integrals are normalised to unity.

Choosing plane-wave background fields with the conventions
\bear
\phi(x) &=& \sum_{i=1}^{N_s} \e^{ip_i\cdot x}  \nonumber\\
\phi_5(x) &=&  \sum_{j=1}^{N_p} \e^{ip_{5j}\cdot x} \nonumber\\
A^\mu (x) &=& \sum_{k=1}^{N_\gamma} \varepsilon_k^\mu \e^{ik_k \cdot x} \nonumber\\
A^\mu_5 (x) &=& \sum_{l=1}^{N_a} \varepsilon_{5l}^\mu \e^{ik_{5l} \cdot x}
\nonumber\\
\label{14-planewave}
\ear
allows us to obtain amplitudes by projecting the effective action onto the multi-linear sector with respect to the particles participating in the scattering process. As in string theory, doing this leads to the following expressions for scalar (`s'), pseudoscalar (`p'), vector (`$\gamma$'), and axialvector (`a') vertex operators that end up being inserted under the worldline path integral:
\bear
V^s_{\rm spin}  [p] &=&
\int_0^Td\tau 
\bigl[z-\sqrt{2}p\cdot\psi -2mi\bigr]
\,\e^{ip\cdot x}
\, ,
\label{defVscal}\\
V^{p}_{\rm spin}  [p_5] &=&
\int_0^T d\tau
\bigl[\hat\gamma_5\big(iz-\sqrt{2}\dot x\cdot \psi+2m\big) -i\sqrt2 Z  \cdot\psi \bigr]
\,\e^{ip_5\cdot x}
\, ,
\label{defVpseudoscal}\\
V^\gamma_{\rm spin}[k,\varepsilon]&=&
\int_0^Td\tau
\bigl[
\varepsilon\cdot \dot x
-i 
\psi\cdot f \cdot\psi
\bigr]
\,\e^{ik\cdot x}
\, ,
\label{defVvector}
\\
V^{a}_{\rm spin}[k_5,\varepsilon_5] &=&
\int_0^Td\tau
\bigl\lbrack \hat\gamma_5\big( i\varepsilon_5\cdot k_5
 + 2\varepsilon_5\cdot\psi\dot x\cdot\psi\big)
-i\,\varepsilon_5\cdot Z
\bigr\rbrack
\,\e^{ik_5\cdot x}\,
\label{defVaxial}
\ear\no
where $f_\mn := k_\mu \varepsilon_\nu - \varepsilon_\mu k_\nu$.
These definitions allow us to represent the one-loop amplitude with $N_s$ scalars, $N_p$ pseudoscalars,  $N_\gamma$ vectors and $N_a$ axialvectors
in the following way:
\bear
\hspace{-2em}\Gamma[\lbrace p_i\rbrace,\lbrace p_{5j}\rbrace,\lbrace k_k,\varepsilon_k\rbrace,
\lbrace k_{5l},\varepsilon_{5l}\rbrace]
&=&
 -\half
 (-ig)^{N_s}
(-ig_5)^{N_p}
(-ie)^{N_\gamma}
(-ie_5)^{N_a}
\nonumber\\
 \hspace{-120pt} &\times&
\int_0^\infty \, 
\frac{dT}{T} 
\e^{-m^2T}
\int Dx
D\psi
Dz
DZ
\e^{-\int_0^Td\tau ( \frac{\dot x^2}{4} + \half \psi\cdot \dot\psi + \frac{z^2}{4}  + \frac{Z^2}{4})}
\nonumber\\
 \hspace{-120pt} &\times&
{\cal P}
\prod_{i=1}^{N_s} V^s_{\rm spin}  [p_i]
\prod_{j=1}^{N_p} V^p_{\rm spin}  [p_{5j}]
\prod_{k=1}^{N_\gamma}V^\gamma_{\rm spin}[k_k,\varepsilon_k]
\prod_{l=1}^{N_a} V^a_{\rm spin}[k_{5l},\varepsilon_{5l}]  \nonumber	
\\
\label{14-ampbig}
\ear\no
where we have now also reinstated the coupling constants. 
The action of the path-ordering operator on a product of $N$ vertex operators can be defined as
\begin{equation}
{\cal P}
(V_1 \cdots V_N) = \sum_{\pi \in S_N} V_{\pi(1)} \cdots V_{\pi(N)}
{\rm \theta}(\tau_{\pi (1)} -\tau_{\pi (2)})
\cdots
{\rm \theta}(\tau_{\pi (N-1)} -\tau_{\pi (N)})
\, ,\hphantom{xx}
\label{14-Pop}
\end{equation}
with $\theta$ the Heaviside function and $S_{N}$ the permutation group on $N$ objects.

\subsection{Summary of evaluation rules}

In the parity-even case, the perturbative evaluation of all the path integrals, including the Grassmann one, proceeds in the same way as in spinor QED.
We can without further ado rewrite the amplitude in terms of Wick contractions:
\bear
\Gamma[\lbrace p_i\rbrace,\lbrace p_{5j}\rbrace,\lbrace k_k,\varepsilon_k\rbrace,
\lbrace k_{5l},\varepsilon_{5l}\rbrace]
&=&
 -2
 (-ig)^{N_s}
(-ig_5)^{N_p}
(-ie)^{N_\gamma}
(-ie_5)^{N_a}
\nonumber\\
&& \hspace{-160pt} \times
\int_0^\infty \frac{dT}{T} \frac{\e^{-m^2T}}{(4\pi T)^{\frac{D}{2}}}
\Bigl\langle
{\cal P}
\prod_{i=1}^{N_s} V^s_{\rm spin}  [p_i]
\prod_{j=1}^{N_p} V^p_{\rm spin}  [p_{5j}]
\prod_{k=1}^{N_\gamma}V^\gamma_{\rm spin}[k_k,\varepsilon_k]
\prod_{l=1}^{N_a}V^a_{\rm spin}[k_{5l},\varepsilon_{5l}] 
\Bigr\rangle
\, .
\nonumber\\
\label{14-ampbigeven}
\ear\no
We leave it as understood that in the vertex operators the original path-integral
variable $x(\tau)$ has to be expanded about the loop centre of mass with the fluctuation variable $q(\tau)$.
The former is integrated out and yields the overall energy-momentum conservation factor 
$(2\pi)^D\delta(\sum_i p_i + \sum_j p_{5j} + \sum_k k_k + \sum_l k_{5l})$, which we suppress,
while the latter is to be Wick-contracted according to 
\eqref{9-wickG}. The spin variable $\psi(\tau)$ is contracted by the familiar \eqref{11-wickgrass2point}, and the 
contraction rules for the auxiliary variables have been given in \eqref{14-wickaux}. 
Although the formalism is still valid in all even dimensions, we have now specialized it to the four-dimensional case, keeping $D$ floating 
only as fas as is necessary for dimensional regularization. 
Thus we have fixed the spin degrees of freedom to be $2^{2}$, but we retain $D = 4 - 2\epsilon$ dimensions for the orbital degrees of freedom.

In the parity-odd case we have to deal with the fermionic zero mode as described in the previous section. 
The formula analogous to \eqref{14-ampbigeven} becomes
\bear
\hspace{-3em}\Gamma[\lbrace p_i\rbrace,\lbrace p_{5j}\rbrace,\lbrace k_k,\varepsilon_k\rbrace,
\lbrace k_{5l},\varepsilon_{5l}\rbrace]
&=&
 - \half
 (-ig)^{N_s}
(-ig_5)^{N_p}
(-ie)^{N_\gamma}
(-ie_5)^{N_a}
\nonumber\\
\hspace{-3em}&& \hspace{-160pt} \times
\int_0^\infty \frac{dT}{T} \frac{\e^{-m^2T}}{(4\pi T)^{\frac{D}{2}}}
\int d^4\psi_0
\Bigl\langle
{\cal P}
\prod_{i=1}^{N_s} V^s_{\rm spin}  [p_i]
\prod_{j=1}^{N_p} V^p_{\rm spin}  [p_{5j}]
\prod_{k=1}^{N_\gamma}V^\gamma_{\rm spin}[k_k,\varepsilon_k]
\prod_{l=1}^{N_a}V^a_{\rm spin}[k_{5l},\varepsilon_{5l}] 
\Bigr\rangle\,,
\nonumber\\
\label{ampbigodd}
\ear\no
where the zero-mode integral has to be done according to \eqref{zeromodeintegral}, and the fermionic Wick contractions using \eqref{xicorr}. 

Introducing some auxiliary Grassmann variables, it would be quite possible to arrive at closed-form master expressions for the 
Wick contractions in \eqref{14-ampbigeven} and \eqref{ampbigodd} (for the pure vector - axial-vector case this has been done in \cite{33}).
However, the procedure is lengthy, and we leave it for future work. Here, we will be satisfied with working out a number of examples. 

\section{Some parity-even amplitudes}
\label{sec-even}

Let us start with the simpler parity-even case. 
At the two-point level, we will calculate the vacuum polarisation amplitudes for scalars and pseudoscalars and redo the axial-vector two-point function presented in \cite{33}) with the Time Slicing compensated vertex operators (the calculation of the vector
one is well-known \cite{strassler1,41}).
As a three-point example, we will also compute the scalar-vector-vector amplitude.

\subsection{Scalar vacuum polarisation} 

The Wick contraction of two scalar vertex operators \eqref{defVscal} yields (with appropriate path ordering)
\bear
\bigl\langle {\cal P}V_{\rm spin}^s[p_1]V_{\rm spin}^s[p_2]\bigr\rangle
=
2\int_0^Td\tau_1 \int_0^{\tau_1}d\tau_2
\Bigl\lbrack 2\delta_{12} +p_1\cdot p_2 G_{F12} -4m^2\Bigr\rbrack \,\e^{G_{12}p_1\cdot p_2}
\, .
\nonumber\\
\ear
As usual in two-point calculations, we use momentum conservation to set $p=p_1=-p_2$, rescale $\tau_i = Tu_i$, and perform the
$T$-integration. Using now also the identity
\bear
\int_0^1du_1\int_0^{u_1}du_2 \, f(u_1-u_2) 
&=&
\half \int_0^1 du f(u) + \half \int_0^1 du u \bigl[f(1-u)-f(u)\bigr]
\nonumber\\
\label{14-idint}
\ear
which holds for any function $f(u)$, we get
\bear
\Gamma^{ss}[p] 
&=& 2\frac{g^2}{(4\pi)^{\Dhalf}}
\biggl\lbrace 2\Gamma\Bigl(1-\Dhalf\Bigr) m^{D-2} - \Gamma\Bigl(2-\Dhalf\Bigr)  \int_0^1 du \frac{p^2+4m^2}{\bigl[m^2+u(1-u)p^2\big]^{2-\Dhalf}}
\biggr\rbrace
\, . \nonumber\\
\label{yukawascalarm}
\ear 
In this form it can be easily checked that it agrees with the result of the Feynman diagram calculation.

\subsection{Pseudoscalar vacuum polarisation}

The Wick contraction of two pseudoscalar vertex operators \eqref{defVpseudoscal} produces
(omitting the subscript `5' on the momenta)
\bear
\hspace{-2em}\bigl\langle {\cal P}V_{\rm spin}^p[p_1]V_{\rm spin}^p[p_2]\bigr\rangle
&=&
2\int_0^Td\tau_1 \int_0^{\tau_1}d\tau_2
\Bigl\lbrack 4m^2 -2\delta_{12} - \bigl(D\ddot G_{12} +  \dot G_{12}^2 p_1\cdot p_2\bigr) G_{F12}\Bigr\rbrack \,\e^{G_{12}p_1\cdot p_2}
\, .
\nonumber\\
\ear
Note that here a minus sign had to be included for the terms involving $\psi$ since, before using $\hat\gamma_5^2=\Eins$, one of
the $\hat\gamma_5$ had to be anticommuted with a $\psi$. As in the calculation of the heat kernel coefficients we must deal with the ambiguity of the product of distributions in the $\ddot{G}_{12}G_{F12} \supset \delta(\tau_{1} - \tau_{2})\textrm{sgn}(\tau_{1} - \tau_{2})$. We have seen that in Time Slicing regularisation, these should be understood as the continuum limits of discrete $\delta$ and sign functions, for which $\textrm{sgn}_{11} = 0$. Hence the $\delta$ function in $\ddot{G}_{12}$ will not contribute and should be omitted. Having resolved this we may fix the sign functions and work in the continuum limit:
 \bear
\hspace{-2em}\bigl\langle {\cal P}V_{\rm spin}^p[p_1]V_{\rm spin}^p[p_2]\bigr\rangle
&=&
2\int_0^Td\tau_1 \int_0^{\tau_1}d\tau_2
\Bigl\lbrack 4m^2 -2\delta_{12} - \bigl(D(\ddot G_{12} - 2\delta_{12}) +  \dot G_{12}^2 p_1\cdot p_2\bigr) \Bigr\rbrack \,\e^{G_{12}p_1\cdot p_2}
\, .
\nonumber\\
\ear
Performing an integration-by-parts on the term involving $\dot G_{12}^2$ turns it into a $\ddot{G}_{12}$, with which the right-hand side can be simplified to
\bear
2\int_0^Td\tau_1 \int_0^{\tau_1}d\tau_2
\Bigl\lbrack \frac{2}{T}(D-1) +4m^2\Bigr\rbrack \,\e^{G_{12}p_1\cdot p_2}
\, .
\nonumber\\
\ear
Proceeding as in the scalar case, we then arrive at
\begin{align}
\Gamma^{pp}[p] 
= 2\frac{g_5^2}{(4\pi)^{\Dhalf}}
\biggl\lbrace 
2(D-1)\Gamma\Bigl(1-\Dhalf\Bigr)  &\int_0^1 du \frac{1}{\Bigl[m^2+u(1-u)p^2\Bigr]^{1-\Dhalf}}
\nonumber\\
 \hspace{50pt} +\,  \Gamma\Bigl(2-\Dhalf\Bigr)  &\int_0^1 du \frac{4m^2}{\Bigl[m^2+u(1-u)p^2\Bigr]^{2-\Dhalf}}
\biggr\rbrace
\, . 
\end{align}
Again this result is in agreement with the Feynman diagram calculation.

\subsection{Axialvector vacuum polarisation}
\label{axialVP}
The axial-vector vacuum polarisation was calculated using naive worldline vertex operators in \cite{33}), that is without the Time Slicing path integral regularisation and counterterms. Here we present the calculation with these amendments, ultimately finding the same result as in \cite{33}.

The contraction of two axial-vector vertex operators provides
\begin{align}
	\big\langle V_{\rm spin}^{a}[k_{1}]V_{\rm spin}^{a}[k_{2}]  \big\rangle = \int_{0}^{T}d\tau_{1} \int_{0}^{T}d\tau_{2} \Big[&\varepsilon_{1} \cdot \varepsilon_{1} \Big((D-1) \ddot{G}_{12} \, \rm{sgn}_{12}\,\rm{sgn}_{21} - 2 \delta_{12}\Big) \nonumber \\
	&+ \varepsilon_{1}\cdot k_{1} \varepsilon_{2}\cdot k_{2} \Big(\dot{G}_{12}\dot{G}_{21}\, \rm{sgn}_{12}\,\rm{sgn}_{21} - 1 \Big) \nonumber \\
	&- \varepsilon_{1}\cdot \varepsilon_{2} k_{1}\cdot k_{2} \dot{G}_{12}\dot{G}_{21}\, \rm{sgn}_{12}\,\rm{sgn}_{21} \Big] \e^{G_{12}k_{1} \cdot k_{2}}
\label{eqVPa}	
\end{align}
where the $\hat{\gamma}_{5}$s squared to the identity. In the first term under the integral, Time Slicing implies that $\delta(\tau_{1} - \tau_{2})\rm{sgn}^{2}(\tau_{1} - \tau_{2}) = 0$ so we arrive at
\begin{align}
	\big\langle V_{\rm spin}^{a}[k_{1}]V_{\rm spin}^{a}[k_{2}]  \big\rangle = \int_{0}^{T}d\tau_{1} \int_{0}^{T}d\tau_{2} \Big[&\varepsilon_{1} \cdot \varepsilon_{1} \Big(\frac{2}{T}(D-1) - 2 \delta_{12}\Big) \nonumber \\
	&+ \varepsilon_{1}\cdot k_{1} \varepsilon_{2}\cdot k_{2} \Big(\dot{G}_{12}\dot{G}_{21}\, \rm{sgn}_{12}\,\rm{sgn}_{21} - 1 \Big) \nonumber \\
	&- \varepsilon_{1}\cdot \varepsilon_{2} k_{1}\cdot k_{2} \dot{G}_{12}\dot{G}_{21}\, \rm{sgn}_{12}\,\rm{sgn}_{21} \Big] \e^{G_{12}k_{1} \cdot k_{2}}
	\label{eqVPaSimple}
\end{align}
This result was also obtained in \cite{33}, but there the vertex operator included an additional prefactor of $\sqrt{D-2}$ in front of the ``$Z$'' term leading to the solo $\delta_{12}$ in (\ref{eqVPaSimple}). However, there the $\delta(\tau_{1} - \tau_{2})\rm{sgn}^{2}(\tau_{1} - \tau_{2})$ was taken to be $\delta(\tau_{1} - \tau_{2})$. Together this provided $\varepsilon_{1}\cdot \varepsilon_{2}\big(2(D-2) - (D-1)\big)\delta_{12} = - 2\varepsilon_{1}\cdot \varepsilon_{2} \delta_{12}$ as obtained above, so the lack of counter term was compensated for by the naive treatment of distributions. 

If we now use momentum conservation to set $k_{1} = k = -k_{2}$ and use the time translation invariance to set $\tau_{2} = 0$ we get a vacuum polarisation tensor by stripping off the polarisation vectors:
\begin{align}
	\Gamma_{a}^{\mu\nu}[k] =  2e_{5}^{2}&\int_{0}^{\infty} \frac{dT}{T} (4\pi T)^{-\frac{D}{2}}\e^{-m^{2}T} \Big\lbrace -2T \delta^{\mu\nu} \nonumber \\
	& + \int_{0}^{T} d\tau \e^{-\tau(1 - \frac{\tau}{T}) k^{2}} \Big[2(D-1)\delta^{\mu\nu} + \big(1 - 2\frac{\tau}{T}\big)^{2}T \big(\delta^{\mu\nu} - k^{\mu}k^{\nu}\big) + Tk^{\mu}k^{\nu}	 \Big] \Big\rbrace\,.
\end{align}
It is now easy to check that in the massless case (a) the first term on the RHS corresponds to a tadpole contribution, which vanishes in dimensional regularisation, and (b) the remaining terms reproduce the vacuum polarisation tensor for the \textit{vector} boson (as to be expected from the chiral symmetry in the parity even case). Equivalence with the Feynman diagram calculation in the massive case was shown in \cite{33}.

\subsection{Scalar-Vector-Vector ($SVV$) amplitude} 

Let us now proceed to a three-point example, the amplitude with one scalar and two photons.
Although the scalar vertex operator \eqref{defVscal} consists of three terms, it is clear that in this
constellation only the one involving $m$ can contribute, since we can Wick contract neither a single
$z$ nor an odd number of $\psi$'s. Thus \eqref{14-ampbigeven} reduces to
\bear
\Gamma^{\gamma\gamma s}
(p;k_1,\varepsilon_1;k_2,\varepsilon_2)
&=&
 -2
 (-ig)
(-ie)^2
\int_0^\infty \frac{dT}{T} \frac{\e^{-m^2T}}{(4\pi T)^2}
\Bigl\langle
V^\gamma_{\rm spin}[k_1,\varepsilon_1]
V^\gamma_{\rm spin}[k_2,\varepsilon_2]
V^s_{\rm spin}  [p]
\Bigr\rangle
\, .
\nonumber\\
\ear\no
Note that we have dropped the path ordering, since this amplitude does not involve terms linear in $\psi$, 
and have also set $D=4$, since this amplitude is finite. 
Wick contraction gives
\bear
\langle V_1^\gamma V_2^\gamma V_3^s \rangle &=& 
(-2mi)
\prod_{i=1}^3
\int_0^Td\tau_i
\Bigl\lbrace
\ddot G_{12} \varepsilon_1\cdot \varepsilon_2 - 
\Bigl(\dot G_{12} \varepsilon_1 \cdot k_2 + \dot G_{13} \varepsilon_1\cdot p \Bigr)
\Bigl(\dot G_{21} \varepsilon_2 \cdot k_1 + \dot G_{23} \varepsilon_2\cdot p \Bigr)
\nonumber\\ &&
- G_{F12}^2 \frac{1}{2} {\rm tr} (f_1f_2) 
\Bigr\rbrace
\, \e^{G_{12}k_1\cdot k_2 + G_{13}k_1\cdot p + G_{23}k_2\cdot p}
\, .
\label{wicksvv}
\ear
We now go on-shell, and fix the helicities of the photons using the standard spinor-helicity formalism (see, e.g., \cite{srednicki-book,elvhua}).
For opposite helicities, using spinor helicity with reference momenta $r_{1,2} = k_{2,1}$ leads to
\bear
\varepsilon_1\cdot k_2 = \varepsilon_2\cdot k_1 = \varepsilon_1\cdot \varepsilon_2 = 0\,, 
\ear
and thus to the vanishing of (\ref{wicksvv}) after eliminating $p$ through momentum conservation.

For equal helicities, we use the same reference momenta, which still leads to the vanishing of
$\varepsilon_1\cdot k_2$ and $\varepsilon_2\cdot k_1$, but now $\varepsilon_1\cdot \varepsilon_2$ survives. 
Using the on-shell relations
\bear
k_1^2 = k_2^2 =0, \, p^2 = - m_s^2, 
\label{onshellscalar}
\ear
as well as momentum conservation we have
\bear
k_1\cdot p = k_2\cdot p = - k_1\cdot k_2 =\frac{m_s^2}{2} \, .
\ear
With these relations, we have
\bear
{\rm tr} (f_1f_2) = m_s^2\, \varepsilon_1\cdot \varepsilon_2 
\ear
and the exponent in (\ref{wicksvv}) simplifies to
\bear
G_{12}k_1\cdot k_2 + G_{13}k_1\cdot p + G_{23}k_2\cdot p =\Bigl(G_{13}+G_{23}-G_{12}\Bigr) \frac{m_s^2}{2} 
\, .
\nonumber
\ear
The prefactor can still be homogeneised, removing the $\ddot G_{12}$ by an integration-by-parts. It then turns into
\bear
&&\Bigl\lbrace \Bigr\rbrace \rightarrow  \varepsilon_1\cdot \varepsilon_2 \, m_s^2 
\biggl\lbrack
\half\Bigl(\dot G_{12}^2- G_{F12}^2\Bigr)
+\fourth\dot G_{12} 
\Bigl(\dot G_{23}- \dot G_{13} \Bigr)
\biggr\rbrack
\, .
\ear
Rescaling $\tau_i = Tu_i$, eliminating the $T$-integral, and introducing the ratio $\tau_s\equiv \frac{m_s^2}{4m^2}$, we have
\bear
\Gamma(p;k_1,\varepsilon_1^\pm;k_2,\varepsilon_2^\pm) &=& - \frac{\alpha}{\pi}  g   \frac{m_s^2}{m} 
\varepsilon_1^\pm \cdot \varepsilon_2^\pm\, I_{SVV}(\tau_s) \, ,\nonumber\\
I_{SVV}(\tau_s) &\equiv& 
\int_0^1 du_1 \int_0^1 du_2 \int_0^1 du_3
\frac
{\half\Bigl(\dot G_{12}^2- G_{F12}^2\Bigr)
+\fourth\dot G_{12} 
\Bigl(\dot G_{23}- \dot G_{13} \Bigr)
}
{1-2\Bigl(G_{13}+G_{23}-G_{12}\Bigr)\tau_s}
\, .
\label{defISVV}
\nonumber\\
\ear
Using the spinor-helicity formalism we can easily calculate
\bear
\varepsilon_1^+ \cdot \varepsilon_2^+ &=& \frac{[12]}{\langle 12\rangle},\quad 
\varepsilon_1^- \cdot \varepsilon_2^- = \frac{\langle 12\rangle}{[12]}  \, .
\label{14-epseps}
\ear
In the integral $I_{SVV}(\tau_s)$, we use the translation invariance to set $u_3=0$, and the symmetry $1\leftrightarrow 2$ to write 
$\int_0^1 du_1 \int_0^1 du_2 = 2 \int_0^1 du_1 \int_0^{u_1} du_2$. Writing out the worldline Green's functions
yields
\bear
I_{SVV}(\tau_s) = 2 \int_0^1 du_1 \int_0^{u_1} du_2 
\frac
{(u_1-u_2)^2-\frac{3}{2}(u_1-u_2)}{1-4u_2(1-u_1)\tau_s}
\, .
\ear
This integral can be calculated by expanding out the denominator as a geometric series, integrating term-by-term, and resumming.
Restricting ourselves now to $0\leq \tau_s \leq 1$, the result can be written as 
\bear
I_{SVV}(\tau_s) = -\frac{1}{2} \frac{\tau_s-(1-\tau_s) \arcsin^2 \sqrt{\tau_s}}{\tau_s^2}
\, .
\label{I(R)fin}
\ear
In the standard model, this amplitude is important for the decay of the Higgs boson into two photons.
Even more important are, however, the decay of the Higgs into two gluons, and the inverse process of Higgs production by gluon fusion. 
Although the extension of the formalism developed in this chapter to the non-abelian case would require more work, for the case at hand it is trivial,
since with only two gluons attached to the fermion loop the only difference to the photon case will be an additional global
colour trace ${\rm tr} \bigl(T^{a_1}T^{a_2}\bigr)$. 
For example, to get the absolute value of the matrix element for Higgs production by gluon fusion with a top-quark loop, we note from
\eqref{14-epseps} that $\abs{\varepsilon_1^+ \cdot \varepsilon_2^+}=\abs{\varepsilon_1^- \cdot \varepsilon_2^-}= 1$
and use that for quarks ${\rm tr} \bigl(T^{a_1}T^{a_2}\bigr) = \half \delta^{a_1a_2}$. Therefore
\bear
|\Gamma(p;k_1,\varepsilon_1^\pm;k_2,\varepsilon_2^\pm)| &=& \delta^{a_1a_2} \frac{\alpha_s}{2\pi}  g_H   m_t^{-1} m_H^2 
 \, I_{SVV}(\tau_H) 
\ear
where now $\tau_H = \frac{m_H^2}{4m_t^2}$ and $g_H = m_t\sqrt{\sqrt{2}G_F}$.  

\section{Some parity-odd amplitudes}
\label{sec-odd}

Let us now consider some parity-odd amplitudes.
\subsection{Pseudoscalar-Vector-Vector ($PVV$) amplitude}
\label{axion}
We start with something simple, namely the three-point amplitude with
one pseudoscalar and two vectors. It is straightforward because we cannot Wick contract terms containing an odd number of $z$, $Z^\mu$
or $\psi^\mu$, so that only the mass term can be used in the pseudoscalar vertex operator \eqref{defVpseudoscal}. 
And the saturation of the zero mode forces us to use the spin part in both vector vertex operators \eqref{defVvector}. 
Thus the general formula \eqref{14-ampbigodd} here reduces to
\bear
\Gamma(p_5,\lbrace k_1,\varepsilon_1\rbrace, \lbrace k_2,\varepsilon_2\rbrace)
&=&
 - \half
(-ig_5)
(-ie)^2
(2m)
(-i)^2
\int_0^\infty \frac{dT}{T} \frac{\e^{-m^2T}}{(4\pi T)^2}
\nonumber\\
&& \hspace{-120pt} \times
\int d^4\psi_0
\Bigl\langle
\int_0^Td\tau_5 \,\e^{ip_5\cdot x_5}
\int_0^Td\tau_1\, \psi_1 \cdot f_1 \cdot \psi_1 \, \e^{ik_1\cdot x_1}
\int_0^Td\tau_2 \,\psi_2 \cdot f_2 \cdot \psi_2 \, \e^{ik_2\cdot x_2}
\Bigr\rangle
\, .
\nonumber\\
\label{14-PVV}
\ear\no
Note that we have set $D=4$, since the amplitude is finite, and we have discarded the path ordering
operator, since no terms linear in $\psi^\mu$ are involved. 
A single global factor of $\hat\gamma_5$ has been taken care of by the periodic boundary conditions. 
Further, the necessity of saturating the fermionic zero-mode integral requires us to always use the $\psi_0^\mu$ after
the split \eqref{splitpsi}. Doing so, and using the zero-mode integral \eqref{zeromodeintegral}, we arrive at 
\bear
\Gamma(p_5,\lbrace k_1,\varepsilon_1\rbrace, \lbrace k_2,\varepsilon_2\rbrace)
&=&
ig_5e^2m
\varepsilon_{\alpha\beta\gamma\delta}f_1^\ab f_2^{\gamma\delta}
\int_0^\infty \frac{dT}{T} \frac{\e^{-m^2T}}{(4\pi T)^2}
\nonumber\\
&&
\Bigl\langle
\int_0^Td\tau_5 \,\e^{ip_5\cdot x_5}
\int_0^Td\tau_1 \, \e^{ik_1\cdot x_1}
\int_0^Td\tau_2  \e^{ik_2\cdot x_2}
\Bigr\rangle
\, .
\nonumber\\
\label{14-PVV2}
\ear\no
Thus the integral that remains to be computed is simply the scalar three-point function.
This object is easy enough to compute off-shell, but let us now go on-shell with the specific application
to axion or pion decay into two photons in mind. 
The on-shell conditions are the same as in the scalar case above, 
just with the scalar mass and momentum replaced by the pseudoscalar ones,
and the integral $I_{SVV}$ gets replaced by
\bear
I_{PVV}(\tau_p) &\equiv& 
\int_0^1 du_1 \int_0^1 du_2 \int_0^1 du_3
\frac
{1}
{1-2\Bigl(G_{13}+G_{23}-G_{12}\Bigr)\tau_p}
=
\frac{\arcsin^2 \sqrt{\tau_p}}{\tau_p}\,,
\label{defIPVV}
\nonumber\\
\ear
where now $\tau_p \equiv \frac{m_p^2}{4m^2}$. The final result becomes
\bear
\Gamma(p_5,\lbrace k_1,\varepsilon_1\rbrace, \lbrace k_2,\varepsilon_2\rbrace)
&=&
\frac{ig_5e^2}{(4\pi)^2} 
\frac  
{
\varepsilon_{\alpha\beta\gamma\delta}f_1^\ab f_2^{\gamma\delta}
}
{m}
\frac{\arcsin^2 \sqrt{\tau_p}}{\tau_p}
\, .
\label{14-PVVfin}
\ear\no
Note that this is imaginary, but only as an artefact of our euclidean conventions.

\subsection{Five pseudoscalar ($PPPPP$) amplitude}

Next, let us look at an example of a parity-odd amplitude with only Yukawa couplings. 
According to \eqref{14-ampbigodd}, any parity-odd amplitude will have a global epsilon tensor
generated by the zero mode of the Grassmann path integral. This tensor must be saturated by four linearly
independent Lorentz vectors, which for a purely scalar amplitude can only be momentum vectors, 
and because of momentum conservation four linearly independent momentum vectors can exist only starting from the
five-point case. Thus we will now have a look at the amplitude with five pseudoscalars, albeit restricted
to the low-energy limit to make a completely explicit calculation feasible.

For this case, \eqref{14-ampbigodd} produces
(dropping the subscript `5' on the momenta)
\bear
\Gamma
[p_1,\ldots,p_5] 
&=&
 - \half
(-ig_5)^5
\int_0^\infty \frac{dT}{T} \frac{\e^{-m^2T}}{(4\pi T)^2}
\int d^4\psi_0\,
\Bigl\langle
{\cal P}
\prod_{j=1}^5\,  V^p_{\rm spin}  [p_{j}]
\Bigr\rangle
\, .
\phantom{xxx}
\label{14-ampbigoddfive}
\ear\no
Looking at the pseudoscalar vertex operator \eqref{defVpseudoscal}, it becomes clear that, to have a chance
of saturating the fermionic zero-mode integral, we have to pick the terms involving $\psi$ in four of the vertex operators, 
say, for legs one to four, and, since a single $z$ or $\xi$ cannot be Wick contracted, the mass term in the fifth. 
Moreover, we cannot use the term involving $\psi$ together with $Z$, since Wick contraction of the latter would produce a pinch that would make 
two momenta merge, thereby reducing the number of linearly independent momenta from four to three. 
Thus the Wick contraction in \eqref{14-ampbigoddfive} can be written as
\bear
\Bigl\langle
{\cal P}
\prod_{j=1}^5 V^p_{\rm spin}  [p_{j}]
\Bigr\rangle
&=& 2m (-\sqrt{2})^4
\Bigl\langle
{\cal P}
\prod_{j=1}^4 
\bigl( \hat\gamma_5 \int_0^Td\tau_j\, \dot x_j \cdot \psi_j \e^{ip_j\cdot x_j}
\bigr)
\hat \gamma_5
\int_0^Td\tau_5 \e^{ip_5\cdot x_5}
\Bigr\rangle
\nonumber\\
&&
+ \, {\rm Perm.} 
\label{wick5point}
\ear
where ``${\rm Perm.}$'' stands for the remaining four possibilities to choose the leg with the mass term.
Further, as in the previous example the necessity of saturating the fermionic zero-mode integral requires us to always use the $\psi_0^\mu$ after
the split \eqref{splitpsi}. We are then ready to perform that integral. Combining \eqref{14-ampbigoddfive} and \eqref{wick5point},
we get
\bear
\Gamma
(p_1,\ldots,p_5) 
&=&
4ig_5^5 m
\varepsilon_{\alpha\beta\gamma\delta}
\int_0^\infty \frac{dT}{T} \frac{\e^{-m^2T}}{(4\pi T)^2}
\int_0^Td\tau_1\int_0^{\tau_1}d\tau_2\int_0^{\tau_2}d\tau_3\int_0^{\tau_3}d\tau_4  
\int_0^T d\tau_5
\nonumber\\
&& \times
\Bigl\langle
\dot x_1^\alpha  \e^{ip_1\cdot x_1}
\dot x_2^\beta  \e^{ip_2\cdot x_2}
\dot x_3^\gamma  \e^{ip_3\cdot x_3}
\dot x_4^\delta  \e^{ip_4\cdot x_4}
\e^{ip_5\cdot x_5}
\Bigr\rangle
+ \, {\rm Perm.} 
\phantom{xxx}
\label{14-ampbigoddfive2}
\ear\no
Note that we have also got rid of the $\hat\gamma_5$ factors, annihilating them in pairs and using the surviving one to 
change the boundary conditions from antiperiodic to periodic. The path ordering has been taken care of by taking
legs 1 to 4 in the standard ordering $\tau_1 \geq \tau_2 \geq \tau_3 \geq \tau_4$ (let us reiterate that
we do not have the option to use translation invariance to set one of the $\tau_i$ equal to zero) and to include in  ``${\rm Perm.}$'' 
a sum over all the other ordered sectors (there is no need to include leg 5 in the ordering, since it did not involve $\psi^\mu$). 
That sum has to be done non-alternatingly, which is contrary to what one would conclude by looking at \eqref{14-ampbigoddfive2} itself,
but clear from \eqref{wick5point} above where the presence of the $\hat\gamma_5$ factors still ensures the correct Bose statistics
with respect to permutations of the pseudoscalars. 

After the removal of the bosonic zero-mode we perform the Wick contractions. 
Contractions between the $\dot q_j^\mu$'s can be discarded since they produce $\delta^\mn$ factors 
Lorentz-contracted with the epsilon tensor. Thus all  $\dot q_j^\mu$'s have to be contracted into the exponential factors, where moreover
terms containing the same $p_j$ twice can be omitted. This leads to
\bear
\Gamma
(p_1,\ldots,p_5)
&=&
4ig_5^5 m
\varepsilon_{\alpha\beta\gamma\delta}
\int_0^\infty \frac{dT}{T} \frac{\e^{-m^2T}}{(4\pi T)^2}
\int_0^Td\tau_1\int_0^{\tau_1}d\tau_2\int_0^{\tau_2}d\tau_3\int_0^{\tau_3}d\tau_4  
\int_0^T d\tau_5
\nonumber\\
& \times &
\prod_{a\ne 1} \dot G_{1a}p^\alpha_a
\prod_{b\ne 2,a} \dot G_{2b}p^\beta_b
\prod_{c\ne 3,a,b} \dot G_{3c}p^\gamma_c
\prod_{d\ne 4,a,b,c} \dot G_{4d}p^\delta_d
\e^{\half \sum_{i,j=1}^5 G_{ij}p_i\cdot p_j}
\nonumber\\
& + & \, {\rm Perm.} 
\phantom{xxx}
\label{14-ampbigoddfive3}
\ear\no
This is still an exact result. However, the exact calculation of five-point integrals is laborious, and we will 
settle for the leading term in the low-energy (``LE'') approximation, being the contribution that is multi-linear also in the 5 particle momenta. This is obtained from \eqref{14-ampbigoddfive3} simply by replacing the
universal exponential factor by unity.

Following this, we then can also discard in the integrand all the terms linear in $p_5$, since their only dependence left on the variable $u_5$ is through a 
single factor of $\dot G_{j5}$, so that they will vanish upon integration in $u_5$. Thus that integration will now simply produce a factor of $T$. 
Performing also the usual rescaling $\tau_i = Tu_i$ for the other four legs, and computing the $T$-integral, we remain with
\bear
\hspace{-1em}\Gamma^{(LE)}
(p_1,\ldots,p_5)
&=&
4ig_5^5 m
\varepsilon_{\alpha\beta\gamma\delta}
p_1^\alpha p_2^\beta p_3^\gamma p_4^\delta 
\frac{2!}{(4\pi)^2m^6}
\int_0^1du_1\int_0^{u_1}du_2\int_0^{u_2}du_3\int_0^{u_3}du_4  
\nonumber\\
 \hspace{-30pt}& \times&
\Bigl\lbrack
\Gd_{12}^2\Gd_{34}^2
+\Gd_{13}^2\Gd_{24}^2
+\Gd_{14}^2\Gd_{23}^2
-2\Gd_{12}\Gd_{23}\Gd_{34}\Gd_{41}
-2\Gd_{12}\Gd_{24}\Gd_{43}\Gd_{31}
\nonumber\\
&& \hspace{2pt}
-2\Gd_{13}\Gd_{32}\Gd_{24}\Gd_{41}
\Bigr\rbrack
 + \, {\rm Perm.} 
\phantom{xxx}
\label{14-ampbigoddfive4}
\ear\no
The total $u$-integral in \eqref{14-ampbigoddfive4} yields a factor of $\frac{1}{120}$.  
Now, another advantage of the low-energy approximation is that
here (but not in general) the $4!$ permutations of legs 1 to 4 all contribute equally. And the same holds true for the final summation over the five possibilities
of choosing the leg with the mass term, since, defining $\varepsilon(p, q, k, l) := \varepsilon_{\alpha\beta \gamma \delta} p^{\alpha}q^{\beta}k^{\gamma}l^{\delta}$, momentum conservation implies that $\varepsilon(p_1,p_2,p_3,p_4) =\varepsilon(p_2,p_3,p_4,p_5) = $ etc. 

This brings us to our final result for the low-energy limit of the amplitude:
\bear
\Gamma^{(LE)}
(p_1,\ldots,p_5)
&=&
4ig_5^5 m
\frac{2!}{(4\pi)^2m^6}
\frac{4!\cdot 5}{120}
\, \varepsilon(p_1,p_2,p_3,p_4)
=
\frac{ig_5^5}{2\pi^2}
\frac
{ \varepsilon(p_1,p_2,p_3,p_4)}{m^5}
\, .
\nonumber\\
\label{14-ampbigoddfivefin}
\ear\no
Once more we leave it to the sceptical reader to verify that the same is obtained with the standard formalism.

\subsection{Vector-vector-axialvector ($VVA$) amplitude and chiral anomaly}

\begin{figure}[htbp]
\begin{center}
\includegraphics[width=0.5\textwidth]{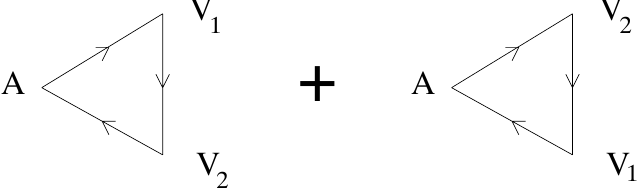}
\caption{Sum of anomalous triangle diagrams in field theory.}
\label{fig-triangle}
\end{center}
\end{figure}

Finally, let us consider the  vector--vector--axial-vector amplitude. In the standard formalism, this one 
is given by the famous ``triangle diagrams'', shown in Fig. \ref{fig-triangle},
that in four-dimensional gauge theory give rise 
to the chiral anomaly and the PCAC (``partially conserved axial current conservation'') relation. 

\vspace{10pt}

For the purpose of deriving that relation, it is sufficient to compute the correlator with the axial-vector replaced by its divergence, $VV\partial\cdot A_{5}$.
In momentum space, the master formula \label{14-ampbigodd} yields for this quantity the following parameter integral,
\bear
\Gamma (\lbrace k_1,\varepsilon_1\rbrace,\lbrace k_2,\varepsilon_2\rbrace , \lbrace k_5,\varepsilon_5 = ik_5\rbrace )
&=&
 - \half
(-ie)^2
(-ie_5)
\nonumber\\
&& \hspace{-160pt} \times
\int_0^\infty \frac{dT}{T} \frac{\e^{-m^2T}}{(4\pi T)^2}
\int d^4\psi_0
\Bigl\langle
V^\gamma_{\rm spin}[k_1,\varepsilon_1] V^\gamma_{\rm spin}[k_2,\varepsilon_2]V^a_{\rm spin}[k_5,ik_5]
\Bigr\rangle
\, .
\nonumber\\
\label{14-ampbigoddanom}
\ear\no
Contrary to our previous examples, there are now various ways of saturating the epsilon-tensor, and the $\xi$-correlator \eqref{xicorr} 
also makes an appearance. Performing the zero-mode integral and Wick contractions, we find
\bear
\Gamma (\lbrace k_1,\varepsilon_1\rbrace,\lbrace k_2,\varepsilon_2\rbrace , \lbrace k_5,\varepsilon_5 = ik_5\rbrace )
&=&
-\frac{i}{2} e^2e_5 \varepsilon_{\alpha\beta\gamma\delta}f_1^\ab f_2^{\gamma\delta}
\int_0^\infty \frac{dT}{T} \frac{\e^{-m^2T}}{(4\pi T)^2}
\prod_{i=1}^3\int_0^Td\tau_i
\nonumber\\&&\hspace{-100pt}
\times \exp\Bigl\lbrack
\Bigl(G_{12} -G_{13}-G_{23}\Bigr)k_1\cdot k_2 
-G_{13}k_1^2 -G_{23}k_2^2
\Bigr\rbrack
\nonumber\\
&&\hspace{-100pt}\times
\biggl\lbrace
(k_1+k_2)^2 +
(\dot G_{12} +\dot G_{23} +\dot G_{31})
(\dot G_{13}-\dot G_{23})
k_1\cdot k_2
-(\ddot G_{13} +\ddot G_{23})
\biggr\rbrace
\, .
\nonumber\\
\label{14-ampbigodd-1}
\ear\no
Here we have also used momentum conservation to eliminate $k_5$. 
%
%
%
%
Removing the second derivatives
$\ddot G_{13}$ ($\ddot G_{23}$) by a
partial integration in $\tau_1$ ($\tau_2$),
the expression in braces turns into
\bear
&&k_1\cdot k_2
\,
\biggl\lbrace
2 -(\dot G_{12}+\dot G_{23}+\dot G_{31})^2
+ \dot G_{12}^2 - \dot G_{13}^2 - \dot G_{23}^2
\biggr\rbrace
+k_1^2 (1-\dot G_{13}^2) + k_2^2 (1-\dot G_{23}^2)
\nonumber\\&&
=
-{4\over T} \Bigl\lbrack
\Bigl(G_{12} -G_{13}-G_{23}\Bigr)k_1\cdot k_2 
-G_{13}k_1^2 -G_{23}k_2^2
\Bigr\rbrack
\label{rewritebraces}
\ear
where in the last step we used the identities 
\bear
\dot G_{ij}^2 = 1 - \frac{4}{T} G_{ij}\, , \quad \dot G_{ij} + \dot G_{jk} + \dot G_{ki} = - G_{Fij}G_{Fjk}G_{Fki} \, .
\ear
Remarkably, the expression in brackets is the same which appears in the exponent of \eqref{14-ampbigodd}!
In the massless case, this leads after the $T$-integration to a total cancellation between numerator and denominator that
trivialises the $u_i$ - integrals, and one obtains without further effort \cite{33}
\bear
\Gamma (\lbrace k_1,\varepsilon_1\rbrace,\lbrace k_2,\varepsilon_2\rbrace , \lbrace k_5,\varepsilon_5 = ik_5\rbrace )\big|_{m=0}
&=&
-\frac{2i}{(4\pi)^2} e^2e_5 \varepsilon_{\alpha\beta\gamma\delta}f_1^\ab f_2^{\gamma\delta}
\, .
\label{14-anomcurr}
\ear\no
In the massive case, the result of the $T$ - integration can be written as
\bear
\Gamma (\lbrace k_1,\varepsilon_1\rbrace,\lbrace k_2,\varepsilon_2\rbrace , \lbrace k_5,\varepsilon_5 = ik_5\rbrace )
&=&
- 2i \frac{e^2e_5}{(4\pi)^2} \varepsilon_{\alpha\beta\gamma\delta}f_1^\ab f_2^{\gamma\delta}
\nonumber\\&&\hspace{-100pt}
\times \prod_{i=1}^3\int_0^1 du_i \,
\frac{ - \Bigl\lbrack  \Bigl(G_{12} -G_{13}-G_{23}\Bigr)k_1\cdot k_2 
-G_{13}k_1^2 -G_{23}k_2^2
\Bigr\rbrack}
{m^2 - \Bigl\lbrack  \Bigl(G_{12} -G_{13}-G_{23}\Bigr)k_1\cdot k_2 
-G_{13}k_1^2 -G_{23}k_2^2
\Bigr\rbrack }
\, .
\nonumber\\
\label{14-anomcurr2}
\ear\no
We then add and subtract $m^2$ in the numerator. The first term produces the previous massless result, 
and the second one can, up to a global factor, be identified with the pseudoscalar-vector-vector amplitude \eqref{14-PVV2}.
Stripping off the polarisation vectors, we get the PCAC relation in the familiar form
\bear
k_5^{\rho}
\langle A_{\mu}A_{\nu}A_{5\rho}\rangle
&=&
-2mi \frac{e_5}{g_5} 
\langle A_{\mu}A_{\nu}\phi_5\rangle
+
{8\over {(4\pi)}^2} e^2 e_5
\,\varepsilon_{\mu\nu\kappa\lambda}k_1^{\kappa}k_2^{\lambda}
\, .
\label{pcac}
\ear
Note that in the present formalism we avoided splitting the amplitude into two diagrams as well as the appearance of UV divergences,
and the anomaly term appears unambiguously at the axialvector current. 
This was to be expected, since the vectors are represented by the usual QED photon vertex operator, 
which turns into the integral of a total derivative when replacing the polarisation vector by the momentum one. 
Nothing like this holds true for the axialvector vertex operator.

\section{Conclusions and Outlook}
\label{sec-conc}

\renewcommand{\theequation}{6.\arabic{equation}}
\setcounter{equation}{0}

In the present manuscript we have presented a novel worldline spinning point particle path integral (\ref{eqLCT}), which for the first time yields the \textit{complete}
effective action, real and imaginary parts, of a Dirac particle simultaneously coupled to external scalar, pseudoscalar, vector and pseudo-vector fields. 
The price which we have to pay for this property is the non-Hermiticity of the
kinetic operator in the exponent, and the presence of some unusual terms with an odd number of Grassmann operators. On the one hand, this requires a specific operator ordering in the worldline action; we use Weyl-ordering which is linked to the Time Slicing regularisation of the path integrals, with its specific rules for the computation of diagrams where a product of distributions arises. Furthermore, a path ordering is also needed when terms with an odd number of Grassmann variables are involved. Unusually for this flat-space context, the use of the Time Slicing regularisation has turned out to require the addition of three counterterms to the
worldline Lagrangian. Curiously, their explicit calculation has led to the elimination of all explicit $D$ - dependence from the worldline Lagrangian, for which we
have presently no explanation. The superrenormalisability of the worldline theory guarantees that no further counterterm additions will be needed to any order
in perturbation theory, which is essential for the practical usefulness of the formalism. 

Our sample calculations clearly show that, at least in perturbation theory, our treatment yields the correct
heat-kernel coefficients and amplitudes, while maintaining the usual advantages of the worldline formalism, such as 
representing amplitudes as a whole rather than through individual Feynman diagrams \cite{15,100,135}, 
and the relative simplicity of accomodating external fields \cite{shaisultanov,18,42,47,ildtor,141}. 
We have also seen that, although the vertex operators
used here generally give rise to a larger number of Wick contractions compared to the more familiar gauge theory amplitudes,
often a large part of these contractions can be discarded already at an early stage of the calculation.

Since our final worldline Lagrangian \eqref{eqLCT} in the pure vector-axialvector case reduces to the one of \cite{29,33}, albeit with the addition of the counterterm \eqref{ctA5}, naturally the question arises whether the existence of this counterterm does not invalidate some of the results of
\cite{29,33,42,47} where this Lagrangian was used. 
Fortunately this is not the case, and a detailed comparison shows that in those calculations the omission of the counterterm was compensated for 
by a different, naive treatment of the delta function contained in $\ddot G_{ij}$. We have shown this for the example of the axialvector vacuum polarisation
in subsection \ref{axialVP}.
However, further study will be required to see whether this equivalence is accidental to low-order calculations, or can be extended to the full perturbation theory. 
Other alternatives to Time Slicing ought to be investigated, too, above all the ``Worldline Dimensional Regularization'' scheme 
 \cite{kleche1,bacova1,bacova2}
which in recent years has become the preferred regularization method for curved-space calculations.

A logical step beyond the present construction would be the extension to the open fermion line, generalising the existing worldline representations for the pure vector background \cite{18,130,131}
(for the case of only the scalar coupling, a worldline treatment of the open fermion line has been given in \cite{alrosc}). The worldline formalism can also be adapted to studying quantum fields on manifolds with boundary and on non-commutative spaces \cite{Bonezzi:2012vr, Ahmadiniaz:2015qaa, Kiem:2001dm, Franchino-Vinas:2018qyk, Ahmadiniaz:2018olx, Manzo:2024gto}, as well as to the treatment of quantum gravity itself~\cite{Bastianelli:2019xhi, Bastianelli:2022pqq, Bastianelli:2023oca, Fecit:2023kah, Fecit:2024jcv} Moreover, the coupling of the particle to non-abelian fields would certainly be another interesting extension, where it is known how suitable auxiliary worldline fields can be introduced to treat the colour structure and avoid a further path-ordering issue \cite{Bastianelli:2013pta, Corradini:2016czo, Edwards:2016acz, Bastianelli:2021rbt, Bastianelli:2021nbs}. 

Alternatively to the worldline formulation, eqs.~(\ref{14-calAW}), (\ref{14-EAHK}) could also be used for writing down a set of second-order
Feynman rules generalising the ones of \cite{33}.

\bigskip

\no
{\bf Acknowledgements:}
JPE is grateful to the hospitality and generosity of the Dipartimento di Scienze Fisiche, Informatiche e Matematiche at the Università degli studi di Modena e Reggio Emilia, where part of this work was carried out, under the ``Visiting professor 2023-2024'' scheme. CS thanks the Dipartimento di Fisica e Astronomia “Augusto Righi”, Università di Bologna, and INFN, Sezione di Bologna, for hospitality.

 \appendix

\section{Heat kernel coefficients}
\label{secApp}
Here we show how the heat kernel coefficients of (\ref{DeWitt}), denoted $a_{n}(x,x)$, can be determined from the effective action \eqref{14-EAHK}, by reusing the well-studied heat-kernel expansion of a non-abelian gauge theory \cite{bles, vassilevich}. In this expansion the first coefficients are given by 

\bea
a_0 &=& \tr \mathbb{I}\nonumber\\
a_1 &=& -\tr\left[\mathfrak a\right]\nonumber\\
a_2 &=& \tr\left[\sbody12 \, {\mathfrak a}^2
- \sbody{1}{12} \, {\mathfrak F}_{\mu\nu} {\mathfrak F}^{\mu\nu} 
\right]\label{a2}\nonumber\\
a_3 &=&  -
\sbody{1}{12} \tr \Big[2{\mathfrak a}^3 +
(D_\mu {\mathfrak a})^2 
 - {\mathfrak a}\,
{\mathfrak F}_{\mu\nu} {\mathfrak F}^{\mu\nu}
-{4\over 15} i 
{\mathfrak F}_{\mu\nu}{\mathfrak F}_{\nu\lambda}{\mathfrak F}_{\lambda\mu}
-{1\over 10}
(D_\lambda{\mathfrak F}_{\mu\nu} )^2 
\Big]\,,
\nonumber\\
\label{a3}
\ear\no
where ${\mathfrak F}_{\mu\nu} \equiv \partial_\mu{\mathfrak A}_\nu - \partial_\nu {\mathfrak A}_\mu +
i[{\mathfrak A}_\mu, {\mathfrak A}_\nu]$,  
$D_\mu {\mathfrak a} \equiv \partial_\mu {\mathfrak a} + i [{\mathfrak A}_\mu,{\mathfrak a}]$, 
and 
$D_\lambda{\mathfrak F}_{\mu\nu} \equiv \partial_\lambda {\mathfrak F}_{\mu\nu}
+  i [{\mathfrak A}_\lambda, {\mathfrak F}_{\mu\nu}] $. 

Let us work this out for the first two coefficients, setting $m=0$ (without loss of generality, since the general result can always be recovered by shifting $\phi \longrightarrow \phi +m$)
but still maintaining the dimensional regularisation. 
%
Taking the spinor dimension to be $2^\frac{D}{2}$ for even $D$
where a nontrivial $\gamma_5$ exists, gives 
\begin{align}
a_0 &= 2^\frac{D}{2}
\nonumber \\
a_1 &= - 2^\frac{D}{2} \Bigl\lbrack 
\phi^2 + (D-1)\phi_5^2 + (D-2)A_5^2
\Bigr\rbrack
\nonumber \\
a_2 
&= 2^\frac{D}{2} \bigg [ 
 \frac16 F^2_{\mu\nu} 
+\frac16 (D-2) (\partial_\mu A_{5\nu})^2 
-\frac13 (\partial A_5)^2 
+\frac16(D-2)(D-4) A_5^4 
\nonumber\\
&
+\frac12(\partial_\mu\phi)^2+\frac12\phi^4
+\frac16 (D-1)(\partial_\mu \phi_5)^2
+\frac16(D-1)(D-3) \phi_5^4
\nonumber\\
&
+ (D-2)(D-3)  \phi_5^2 A_5^2
+(D-2)\phi^2 A_5^2
+ (D-3) \phi^2\phi_5^2
\nonumber\\
&
-2(D-2) \phi_5 A_5^\mu \partial_\mu \phi
-2\phi\phi_5 \partial \cdot A_5 
\bigg ] \,.
\label{computea12}
\end{align}

For $D = 2$ we find
\begin{align}
a_{0} &= 2\\
a_{1} &= -2\Big[ \phi^{2} + \phi_{5}^{2}\Big]\\
a_{2} &= 2\bigg [ 
 \frac{1}{6} F^2_{\mu\nu} 
- \frac13 (\partial A_5)^2 
+\frac12(\partial_\mu\phi)^2+\frac12\phi^4
\nonumber\\
&
+\frac{1}{6} (\partial_\mu \phi_5)^2
-\frac{1}{6} \phi_5^4  
-
\phi^2\phi_5^2 -2\phi\phi_5 \partial \cdot A_5 \bigg ]\,.
\end{align}
Similarly, in $D = 4$ these become
\begin{align}
a_{0} &= 4\\
a_{1} &= -4\big[ \phi^{2} + 3\phi_{5}^{2} + 2A_{5}^{2}\Big]\\
 a_{2} &= 
4\bigg [ 
 \frac{1}{6} F^2_{\mu\nu} 
+\frac13 (\partial_\mu A_{5\nu})^2 
- \frac13 (\partial A_5)^2 
+\frac12(\partial_\mu\phi)^2+\frac12\phi^4
\nonumber\\
&
+\frac12 (\partial_\mu \phi_5)^2
+\frac12 \phi_5^4 + 2\,  \phi_5^2 A_5^2  
+2\,  \phi^2 A_5^2 
+
\phi^2\phi_5^2
\nonumber\\
&
-4\phi_5A_5^\mu \partial_\mu \phi
-2\phi\phi_5 \partial \cdot A_5 \bigg ]
\, .
\end{align}

Up to total derivatives, the last term can also be written as 
\begin{align}
a_2&= 4\bigg [ 
 \frac{1}{6} F^2_{\mu\nu} 
+\frac16 F^2_{5\, \mu\nu} 
+\frac12(\partial_\mu\phi)^2+\frac12\phi^4
\nonumber\\
&
+\frac12 (\partial_\mu \phi_5)^2
+\frac12 \phi_5^4 
+2\,  \phi_5^2 A_5^2  
+2\,  \phi^2 A_5^2 
+
\phi^2\phi_5^2
\nonumber\\
& + 2 A_5^\mu ( \phi \partial_\mu \phi_5 - \phi_5 \partial_\mu \phi)  \bigg] \,.
\end{align}

Note that certain coefficients show a continuous phase symmetry with respect to a doublet, $\Phi := \begin{pmatrix}\phi \\ \phi_{5}\end{pmatrix}$, under $\Phi \rightarrow R(\alpha)\Phi$ with $R(\alpha)$ an $\textrm{SO}(2)$ rotation by angle $\alpha$. This is a symmetry of the Dirac theory (\ref{Sspinhalfmink}) as it can -- in the massless limit -- be absorbed by a (global) chiral transformation of $\psi \to \e^{i\alpha \gamma_5} \psi$. It changes the effective action, (\ref{defEAeuc-annchiral}), by a constant if we absorb the change by defining a new, equivalent set of gamma matrices, $\Gamma^{\mu} = \e^{-2i\alpha \gamma_{5}}\gamma^{\mu}$ (with $\Gamma_{5} = \gamma_{5}$). 

This symmetry carries over to the amplitudes, as can, in the standard formalism, be easily verified diagrammatically using the fact that the
Dirac propagator in this limit anticommutes with $\gamma_5$. 
However, since the Seeley-DeWitt expansion is a large mass expansion, here the only coefficients where we can expect it to be respected are those in $\Gamma[\phi, \phi_{5}, A, A_{5}]$ that are independent of mass (and therefore correspond to logarithmic divergences -- see (\ref{14-EAHK})). These are precisely $a_{1}$ for $D = 2$ and $a_{2}$ for $D = 4$, where this $SO(2)$ symmetry is manifest:
\begin{align}
	a_{1} &\overset{D=2}{=} -2 \big|\Phi\big|^{2}\\
	a_{2}  &\overset{D=4}{=} 4\Big[\frac{1}{2} \big|\partial_{\mu}\Phi\big|^{2} + \frac{1}{2}\big|\Phi\big|^{4}  + 2A_{5}^{2}\big|\Phi\big|^{2} + 2A_{5}^{\mu}\epsilon_{ij}\Phi^{i}\partial_{\mu}\Phi^{j}\Big]\,.
\end{align}

\end{document}